\documentclass[cameraready]{Interspeech}

\usepackage{cite}
\usepackage{amsmath,amssymb,amsfonts}
\usepackage{algorithmic}
\usepackage{graphicx}
\usepackage{algorithm,algorithmic}
\usepackage{hyperref}
\hypersetup{hidelinks=true}
\usepackage{textcomp}

\usepackage{comment}
\usepackage{booktabs}
\usepackage[table]{xcolor} 
\usepackage{colortbl} 

\definecolor{low}{RGB}{198, 239, 206}    
\definecolor{medium}{RGB}{255, 235, 156} 
\definecolor{high}{RGB}{255, 199, 206}   

\usepackage{cite}
\usepackage{amsmath,amssymb,amsfonts}
\usepackage{algorithmic}
\usepackage{graphicx}
\usepackage{textcomp}
\usepackage{xcolor}
\usepackage{subfigure}
\usepackage{caption, subcaption}
\usepackage{amsmath,graphicx}
\usepackage{textcomp}
\usepackage{amsmath,amssymb,amsfonts}
\usepackage{mathtools}
\usepackage{amsmath,graphicx, multirow, hyperref}
\usepackage{subcaption, subfigure}
\usepackage{graphicx,subfigure}

\newcommand{\cognospeak}{CognoMemory}
\newcommand{\cognospeakOLD}{CognoSpeak}
\newcommand{\bart}{BART}

\newcommand{\roberta}{RoBERTa}
\newcommand{\distilbert}{DistilBERT}

\newcommand{\adress}{ADReSS}

\newcommand{\tfidf}{TF-IDF}

\newcommand{\Whisper}{Whisper}
\newcommand{\WV}{Wav2Vec 2.0}
\newcommand{\Nemo}{NeMo}

\newcommand{\CognoG}{Cogno\_G}
\newcommand{\CognoZ}{Cogno\_Z}

\usepackage{comment}

\title{Improved automatic speech-based cognitive assessment using neuropsychologically calibrated diagnostic labels}

\author[affiliation={1}]{Madhurananda}{Pahar}
\author[affiliation={2}]{Caitlin}{Illingworth}
\author[affiliation={3}]{Adam}{Hampshire}
\author[affiliation={3}]{Peter J}{Hellyer}
\author[affiliation={4}]{Agata J}{Hampshire}
\author[affiliation={2}]{Daniel}{Blackburn}
\author[affiliation={2}]{Annalena}{Venneri}
\author[affiliation={1}]{Heidi}{Christensen}

\address{
    $^1$ School of Computer Science, The University of Sheffield, Sheffield, S1 4DP, UK \\
    $^2$ Sheffield Institute for Translational Neuroscience (SITraN), The University of Sheffield, Sheffield, S10 2HQ, UK \\
    $^3$ Centre for Neuroimaging Sciences, IoPPN, King's College London, London, UK \\
    $^4$ Wolfson Sensory, Pain and Regeneration Centre (SPaRC ), King’s College London, UK
}

\email{m.pahar@sheffield.ac.uk, c.illingworth@sheffield.ac.uk, adam.hampshire@kcl.ac.uk, peter.hellyer@cognitron.co.uk, agata@cognitron.co.uk, d.blackburn@sheffield.ac.uk, a.venneri@sheffield.ac.uk, heidi.christensen@sheffield.ac.uk}


\begin{document}

\maketitle

\begin{abstract}

    Although automatic detection of cognitive impairment from conversational speech is a promising way, it becomes less accurate, as large real-world datasets often contain unreliable diagnostic labels due to self-referral and limited clinical verification, and here we investigated whether neuropsychologically calibrated diagnostic relabeling improves it. 
    We analysed 138 hours of conversational speech from 720 participants, recalibrating unreliable diagnostic labels using attention, executive function, and memory performance via external clinical thresholds, Gaussian, and normative z-score calibration. Transformer-based classifiers were trained using both original and recalibrated labels and evaluated internally on 150 clinically verified participants and externally on Pitt, \adress{} and PROCESS-2 datasets. 
    Memory- and MoCA-based z-score calibration produced the greatest improvements. Using our internal dataset, $F_1$ increased from $0.70$ to $0.83$ ($+13\%$) and from $0.54$ to $0.66$ ($+12\%$) for binary and multiclass classification, respectively.
    For external validation, the increases are $+4\%$, $+5\%$ and $+9\%$ on Pitt, \adress{} and PROCESS-2 corpus. 
    Neuropsychologically calibrated relabeling substantially improves speech-based cognitive assessment, highlighting label quality, rather than model architecture, as a critical bottleneck in clinical AI for dementia screening.

\end{abstract}


\section{Introduction}
\label{sec:intro}


Cognitive and memory difficulties can arise from a range of factors, including fatigue, stress, and illness, and tend to become more pronounced with age. When these deficits persist over several months, they may indicate mild cognitive impairment (MCI) \cite{davis2018estimating}, a condition characterised by deficits in memory, learning, reasoning, attention, language, and motivation. Individuals with MCI often describe experiencing a reduced capacity for clear thinking \cite{rosenberg2013association}. 
Approximately half of those diagnosed with MCI eventually progress to dementia, a progressive neurodegenerative condition, most commonly resulting from Alzheimer’s disease (AD) \cite{prestia2013prediction}. 
Other forms of dementia, including vascular and frontotemporal dementia, frequently disrupt behaviour, language, and cognitive functions such as perception and executive reasoning. 
To date, no curative therapies exist for these conditions, only those that slow the rate of decline \cite{knopman2003essentials, thabtah2020correlation, hendrie1998epidemiology}. 


Early identification of dementia provides significant benefits, including the opportunity for timely interventions that may delay progression to advanced stages \cite{robin2021using}. However, diagnosis is often challenging, requiring comprehensive evaluations by neurological specialists, longitudinal cognitive assessments maintained by healthcare providers \cite{shi2023speech}, and complex, expensive, and time-consuming clinical procedures such as magnetic resonance imaging (MRI), standardised cognitive testing, and cerebrospinal fluid analysis \cite{yang2022deep}. These traditional approaches are impractical for large-scale screening of cognitive decline \cite{mckhann2011diagnosis} and contribute to underdiagnosis, leaving many individuals without timely treatment. Consequently, there is a growing need for remote and intelligent technologies capable of supporting healthcare services in delivering accurate and timely assessments \cite{gauthier2021world}, including approaches that leverage speech, behaviour, and other non-invasive data sources for scalable early detection.


For this purpose, we have previously designed \cognospeak{} (formerly \cognospeakOLD{\cite{pahar2025cognospeak}}) to accelerate the identification of early cognitive decline by engaging the participant in a conversation, which places demands on multiple cognitive domains including memory, language and attention, and subsequently carefully analysing the answers, as described in \cite{pahar2025cognospeak, pahar2025mutlimodalfusion}. 
This remote, large-scale data collection, either at patients' homes, at clinics or in community centres, allows us to capture real-world conversational speech, and a similar automatic system will save time and financial resources for health service providers across the world, such as the NHS in the UK. 
Although the \cognospeak{} system is primarily used to collect data from primary or secondary care memory clinics in the UK (clinical source), it also recruits nationally through volunteer websites and social media (non-clinical source), as it is open to the public. 
The diagnostic labels of the participants from non-clinical sources is provided by the participants themselves and is therefore not verified by a clinician involved in data collection. 
This introduces a significant source of label uncertainty for downstream machine learning tasks, as the provided annotations may lack clinical accuracy. 
Participants may rely on outdated historical diagnoses (such as an individual previously diagnosed with MCI who has since progressed to dementia), or they may inadvertently provide an inaccurate self-report due to a lack of awareness in terms of their own cognitive status. 
Given the limitations of collecting data through online databases and to address uncertainty in unverified diagnostic annotations, we explore ways to improve diagnostic label accuracy by adopting a neuropsychologically calibrated label reassignment strategy in which diagnostic categories are inferred from objective cognitive performance measures and subsequently validated against clinically established ground-truth diagnoses.


This study proposes a clinically grounded, threshold-based neuropsychological relabelling framework that replaces self-reported diagnostic labels with neuropsychologically stratified categories derived from measured performance. 
Specifically, we develop and evaluate relabelling approaches using three complementary thresholds: (i) clinical reference thresholds from an external study (FLAME) \cite{brooker2020flame}, (ii) Gaussian-based decision boundaries estimated from trusted clinical subsets of \cognospeak{}, and (iii) normative deviation thresholds based on healthy-control z-scores. The reassigned labels are subsequently validated against internal (clinical ground-truth diagnoses from \cognospeak{}) and three external datasets (Pitt \cite{becker1994natural}, \adress{} \cite{luz2020alzheimer}), and PROCESS-2 \cite{pahar2026PROCESS2} through extensive speech-based classification experiments. 
Our results demonstrate that improving diagnostic label quality produces larger performance gains than modifying model architecture, highlighting data curation as a primary bottleneck in automatic cognitive decline detection. 
The novelty of this work lies in systematically investigating whether cognitive performance can be used to improve diagnostic supervision for speech-based cognitive assessment. 
Thus, we pose the following research questions. 
First, how do the neuropsychological profiles of participants recruited through \cognospeak{} compare with those reported in the FLAME cohort, the only closely comparable reference dataset available?
Second, whether relabelling participants with unconfirmed diagnoses using objective cognitive scores improves automatic speech-based classification performance and whether any observed improvements are statistically significant.
Third, whether incorporating MoCA scores alongside neuropsychological measures further enhances classification accuracy.
Finally, which one among three complementary relabelling strategies, FLAME-derived thresholds, Gaussian-calibrated thresholds, and normative z-score thresholds, provides the most reliable supervisory signal; and how do the relative importance of attention, executive function, and memory domains for estimating clinically meaningful diagnostic boundaries assessed.

The remainder of the paper answers these questions. 
Section~\ref{section:data} describes participant recruitment and dataset construction, demonstrating strong agreement between our cohort and the FLAME dataset, particularly in memory performance, supporting cross-dataset generalisability and external validity of the proposed cognitive representation. The same section further shows that, while small domain-dependent differences appear in attention and executive function across clinical and non-clinical sources, memory scores remain stable, suggesting robustness to data-source heterogeneity, answering the first research question. 
Section~\ref{sec:results} evaluates the impact of neuropsychological relabelling on downstream classification, and answers the rest of the research questions. 
Across both binary and multiclass diagnostic tasks, models using the memory domain and z-score based relabelling consistently achieve the highest $F_1$-scores.
Improvements over the original labels are statistically significant in binary classification (up to +0.13 $F_1$), and they are the most pronounced in the more challenging three-class setting (up to +0.12), reflecting increased diagnostic relabelling accuracy for the cognitively impaired cohort (dementia and MCI) for the internal validation dataset.
The same relabelling strategy has also performed the best among others for three external datasets: Pitt, \adress{} and PROCESS-2, each with an improvement of 4\%, 5\% and 9\%. 
Adding MoCA into the diagnostic relabelling strategy also increases performance up to 8 and 12\% for binary and multiclass classification, highlighting the importance of MoCA. 
Finally, Section~\ref{sec:discussion} provides interpretability analysis demonstrating that neuropsychologically informed labels reveal distinct semantic retrieval patterns in participant speech. \tfidf{} analyses link improved classification performance to interpretable linguistic biomarkers of cognitive decline, supporting the clinical relevance of the proposed relabelling framework. 
Taken together, this study establishes a principled methodology for transforming weakly labelled cognitive datasets into clinically meaningful resources, thereby advancing scalable, speech-based dementia screening and remote cognitive monitoring.

\begin{figure*}[!ht]
  \centering 
  \includegraphics[width=\linewidth]{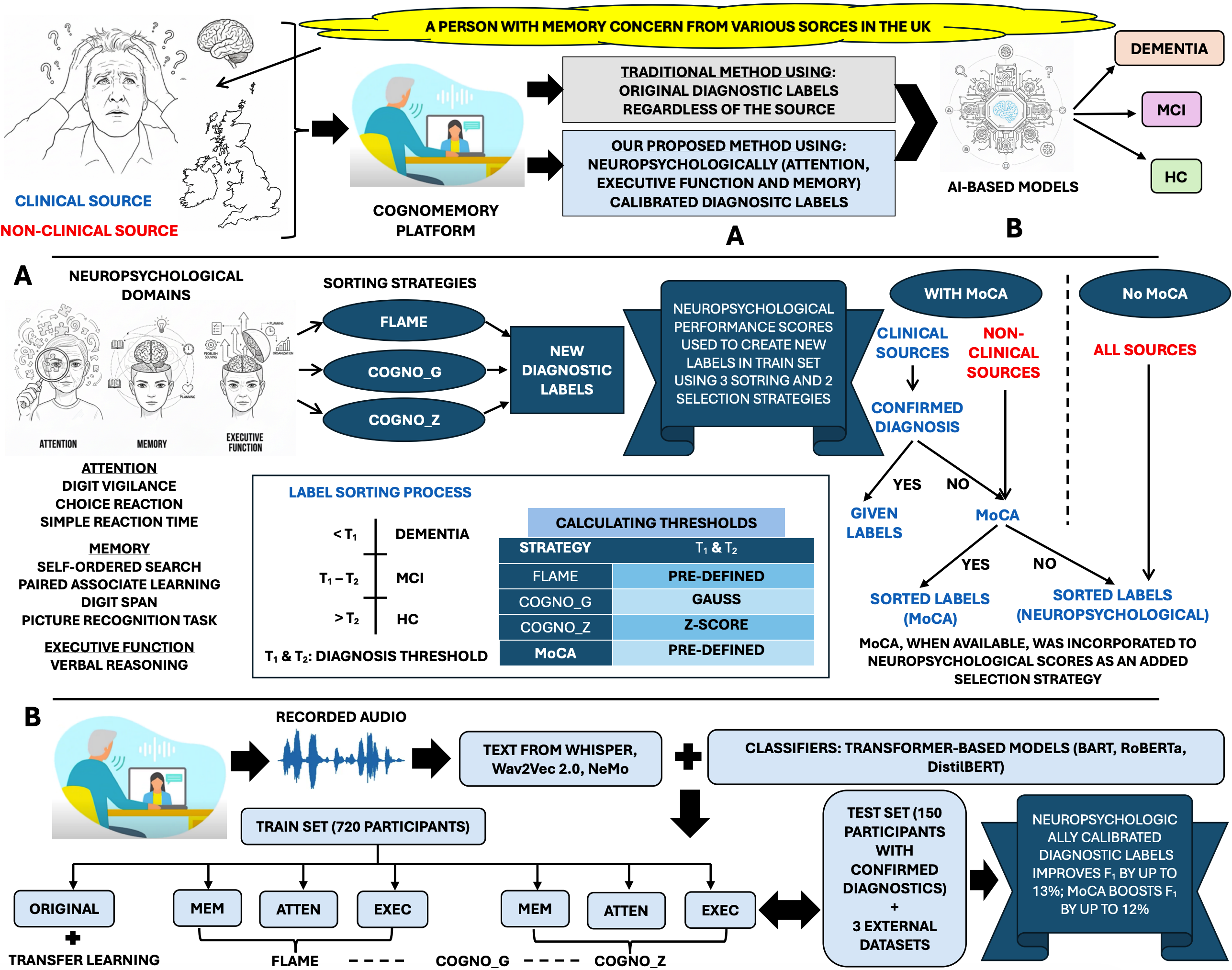} 
  \caption{
    \textbf{Overview of the proposed improved neuropsychological performance-based diagnostic relabelling and speech-based classification framework.}
    Participants with memory concerns were recruited across the United Kingdom (UK) from both clinically confirmed and non-clinical community sources and completed remote cognitive assessments using the \cognospeak{} platform. 
    \textbf{(A) Neuropsychological label reassignment.} 
    Instead of directly using inherited diagnostic labels, neuropsychological performances across attention, executive function and memory domains were analysed to generate refined diagnostic labels. 
    Three complementary threshold-based sorting strategies were applied: external FLAME \cite{brooker2020flame} thresholds, data-driven Gaussian calibration (\CognoG{}), and normative z-score deviation thresholds (\CognoZ{}). 
    Two selection strategies were evaluated: `with MoCA' integrates clinically validated diagnoses and MoCA-derived priors when available, whereas `No MoCA' derives labels solely from neuropsychological performance, enabling large-scale weakly supervised dataset curation. 
    Threshold values are noted as $T_1$ and $T_2$, which are used to relabel participants as dementia, mild cognitive impairment (MCI), and healthy control (HC) groups.
    \textbf{(B) Speech-based model training and evaluation.}
    Texts from speech recordings collected during cognitive testing were generated using three ASRs, such as \Whisper{}, \WV{} and \Nemo{}. Classification was performed using those texts by three transformer-based models, like \bart{}, \roberta{} and \distilbert{}. 
    Models were trained using neuropsychologically calibrated diagnostic labels and evaluated against an independent test set with ground-truth clinical diagnoses, demonstrating improved performance compared to the original label usage. 
    }
  \label{fig:summary_paper}
\end{figure*}



\section{Previous work}
\label{sec:previouswork}

\subsection{Neuropsychological Score–Based Diagnostic Stratification}

Obtaining reliable diagnostic labels remains a fundamental challenge in computational studies of cognitive impairment.
Clinical diagnosis of dementia and MCI requires expert interpretation of neuroimaging, biomarkers, informant report, collateral history, and cognitive assessments \cite{NICE2018Dementia}, making large-scale clinically verified datasets difficult to obtain.
Consequently, many neuropsychological and epidemiological studies operationalise diagnosis using cognitive screening scores rather than clinically confirmed diagnostic labels. 

Traditional neuropsychology defines impairment relative to normative populations using standardised deviation thresholds, typically classifying MCI at approximately $1$--$1.5$ standard deviations below healthy performance and dementia at more severe deviations \cite{petersen1999mild, albert2011diagnosis}.
Large cohort studies such as the Alzheimer's Disease Neuroimaging Initiative (ADNI) employ neuropsychologically anchored cut-offs derived from clinically characterised reference populations to stabilise diagnostic categorisation \cite{jack2008alzheimer}.

Data-driven calibration of diagnostic thresholds has previously been explored for improving cognitive impairment characterisation under uncertain clinical supervision \cite{jak2009quantification}. 
Gaussian and distribution-based modelling approaches enable diagnostic boundaries to be estimated directly from population statistics when expert annotations are limited or heterogeneous \cite{reynolds2009gaussian, ashman2000mixture}. 
Such approaches additionally treat cognitive decline as a continuous spectrum rather than strictly discrete categories, improving robustness to inter-cohort variability and intermediate disease states such as MCI \cite{petersen2016mild, jack2018nia}.

\subsection{Speech-Based Detection of Cognitive Impairment}

A person's speech and language provide sensitive markers of neurodegeneration, motivating extensive research into automated dementia detection from speech recordings.
Early studies using the DementiaBank Pitt Corpus showed that lexical diversity, pause structure, and syntactic complexity correlate strongly with cognitive decline \cite{becker1994natural}.
Subsequent work demonstrated that acoustic and linguistic features extracted from spontaneous speech can distinguish dementia from healthy ageing with high accuracy \cite{konig2015automatic, pan2021using, kumar2022dementia}.

More recent approaches employ deep learning and large-scale speech modelling to predict cognitive status or disease progression directly from audio recordings \cite{mirheidari2018detecting, mirheidari2023identifying, syed2021tackling, luz2020alzheimer, young2025can}.
Importantly, several studies report that speech-derived biomarkers reflect underlying cognitive domains such as memory retrieval and executive control rather than merely capturing superficial linguistic differences \cite{robin2021using}.
This shift toward cognitively interpretable modelling highlights the importance of aligning machine learning pipelines with neuropsychological theory.

\subsection{Learning Under Noisy or Weak Clinical Labels}


Several studies have addressed the problem of unreliable supervision in medical machine learning, where real-world clinical datasets frequently contain incomplete, weak, or inconsistent annotations \cite{karimi2020deep, northcutt2021confident}. 
Such limitations have motivated the development of weak-supervision and label-refinement frameworks. 
Methods such as probabilistic label modelling and data programming demonstrate that combining imperfect signals can produce higher-quality supervisory labels than raw annotations alone \cite{ratner2019snorkel}. 
Similarly, learning with noisy labels has been shown to benefit substantially from label correction or relabelling strategies guided by domain knowledge \cite{frenay2013classification, goldberger2017training}.



\subsection{Position of the Present Work}

Prior dementia-detection studies typically either rely exclusively on clinically confirmed labels from limited-size cohorts or train directly on noisy diagnostic annotations without systematic refinement. To the best of our knowledge, this is the first work to explicitly reconstruct diagnostic labels using neuropsychologically stratified thresholds and subsequently validate them against independent ground-truth clinical diagnoses. 
Unlike prior speech-based dementia studies that rely directly on inherited diagnostic annotations, we reconstruct labels by combining clinically anchored diagnostic stratification strategies across attention, executive function, and memory domains for speech-based cognitive decline modelling. Rather than treating unconfirmed labels as fixed supervision, we derive new diagnostic boundaries from normative z-scores of domain-level neuropsychological performance, enabling improved automatic detection of cognitive decline. 




\section{Data}
\label{section:data}

\subsection{\cognospeak{} data collection along with neuropsychological stratification}

\cognospeak{} is an online, innovative AI tool which engages with the participant by asking 14 memory-probing, clinically effective questions, including motivation, quantifying memory, concern assessments, long- and short-term memory tests, semantic \& phonemic fluency tests, cognitive functioning, cookie-theft and complex picture description tasks in a fixed order, to automatically analyse the possible signs of cognitive decline. 
The questions and the data collection procedure have been largely explained in our previous studies \cite{pahar2025cognospeak, pahar2025mutlimodalfusion}. 

For this study, we implemented an online platform \footnote{\url{https://cognospeak.cognitron.co.uk/}} to collect neuropsychological performance across three domains, such as attention accuracy, executive function and memory. 
Participants are asked to complete 8 different tasks which correspond to three neuropsychological domains, as described in Table 1 of the FLAME study \cite{brooker2020flame}.
Each scores are summed up for each category. 

\subsection{\cognospeak{} diagnosis labels and data description}

To date, more than 2000 participants have been recruited through the \cognospeak{} platform, of whom 720 completed the extended neuropsychological performance assessment used in this study. The distribution of diagnostic labels is summarised in Table~\ref{table:dataset}. The largest proportion of participants consisted of healthy controls (HC), while the dataset additionally included 25 participants with dementia and 90 with mild cognitive impairment (MCI). 
Participants were recruited from both clinical and non-clinical sources across the United Kingdom. Clinical recruitment pathways included primary and secondary care services, whereas non-clinical recruitment involved sources such as Join Dementia Research, Great Minds, self-referrals, social media, word of mouth, and unknown community pathways. 
In this study, multilingual participants, predominantly representing ethnic minority communities including South Asian, Chinese, and African populations in the UK, were excluded because our recent work demonstrated that current AI systems are not yet sufficiently reliable for detecting cognitive decline in multilingual speech \cite{pahar2026can, illingworth2026assessing}.

Among the dementia cohort, 6 participants originated from clinically confirmed sources and 19 from non-clinical recruitment pathways. Similarly, the MCI cohort consisted of 21 clinically recruited and 69 non-clinically recruited participants. In contrast, the HC cohort was predominantly community-based, consisting of 9 clinically recruited and 456 non-clinically recruited participants. Overall, approximately 80\% of the cognitively impaired participants originated from clinically supervised recruitment streams, whereas the HC group was largely derived from community recruitment. 
The dementia cohort had a mean age of 71.76 (8.93) years, whereas the MCI and HC groups were younger, with mean ages of 69.16 (7.61) and 65.95 (8.28) years, respectively. Female participants were more prevalent than male participants in all groups except MCI, comprising approximately 60\% of the total cohort.

A total of 138 hours of speech audio spanning all 14 cognitively effective questions was collected from the 720 participants, including 140 participants with unconfirmed diagnoses \cite{pahar2025mutlimodalfusion}, whose data were subsequently incorporated into the proposed neuropsychological relabelling framework for classifier training, explained in Section \ref{subsec:labelReassign}. 
Automatic speech recognition (ASR) systems, including Whisper-medium \cite{radford2022whisper}, \WV{} \cite{baevski2020wav2vec}, and NVIDIA \Nemo{} (NVIDIA Neural Modules 1.21.0) \cite{kuchaiev2019nemo} were employed to generate transcripts from recorded speech, producing approximately 900.21, 946.56, and 908.05 thousand words, respectively.

A handcrafted independent test set of 150 participants was additionally prepared. Unlike the training cohort, all participants in the test set were recruited exclusively through clinical pathways and therefore possessed clinically confirmed ground-truth diagnostic labels. The test set size was selected to represent approximately 20\% of the training cohort size (720 participants), corresponding to a standard 5-fold cross-validation evaluation scheme commonly used in machine learning studies \cite{racz2021effect, pahar2024nanocnn, pahar2026ccman}. 
The test set consisted of 23 dementia, 50 MCI, and 77 HC participants, with balanced male and female representation. Similar demographic trends were observed to those in the training cohort, with dementia participants being older on average than MCI and HC participants, although the overall test cohort was slightly older than the training cohort.

In total, 870 participants were included across training and testing phases. A subset of participants additionally had Montreal Cognitive Assessment (MoCA) scores available, which were incorporated into the clinically informed sorting strategies illustrated in Fig.~\ref{fig:summary_paper}.
Because the overall cohort size remained limited, strict demographic and diagnostic balancing across age, gender, and cognitive groups was not feasible without substantially reducing the number of available participants; this limitation is discussed further in Section~\ref{subsec:limitations}.

\begin{table*}[h]
    \centering
    \footnotesize
    \setlength{\tabcolsep}{4pt} 
    \renewcommand\arraystretch{1.3}
    \caption{\textbf{Dataset description.} A total of 720 participants who completed both \cognospeak{} and the neuropsychological assessments were used in training. Another 150 participants who participated in only \cognospeak{} with ground truth diagnostic labels were used for testing. The gender distribution excludes 11 undisclosed participants. The number of words generated by Whisper, \WV{}, and \Nemo{} is listed as 1,000. Clinical and non-clinical are noted as \textit{Src(C)} and \textit{Src(NC)}. }
    \label{table:dataset}
    \begin{tabular}{c | c | c | c | c | c | c | c | c | c | c | c | c | c }
    
    \toprule
    \textbf{Evaluation} & \textbf{Group} & \textbf{Diagnosis} & \textbf{\textit{N}} & \textbf{\textit{Src(C)}} & \textbf{\textit{Src(NC)}} & \textbf{Age} & \textbf{M} & \textbf{F} & \textbf{Audio (hr)} & \textbf{Whisper} & \textbf{Wav2Vec2} & \textbf{\Nemo{}} & \textbf{MoCA} \\ 
    
    \midrule
    \multirow{4.5}{*}{Train} & \multirow{2.5}{*}{Case} & Dementia  & 25 & 6 & 19 & 71.76$\pm$8.93  & 18  & 7 & 5.76 & 31.53 & 34.19 & 32.99 & 22 \\ 
    \cmidrule{3-14}
     &  & MCI & 90 & 21 & 69 & 69.16$\pm$7.61  & 45  & 44 & 19.52 & 121.33 & 126.53 & 120.46 & 69 \\
    \cmidrule{2-14}
     & Control & HC & 465 & 9 & 456 & 65.95$\pm$8.28 & 172  & 286 & 85.6 & 564.89 & 593.68 & 571.92 & 8 \\
    \cmidrule{2-14}
     & OTHER & Unconfirmed & 140 & 0 & 140 & 66.15$\pm$8.43 & 51 & 86 & 27.13 & 182.46 & 192.17 & 182.68 & 10 \\

    \cmidrule{1-14}
    \textbf{SUBTOTAL} & --- & --- & \textbf{720} &\textbf{36} & \textbf{684} & \textbf{66.58$\pm$8.38} & \textbf{286} & \textbf{423} & \textbf{137.99} & \textbf{900.21} & \textbf{946.56} & \textbf{908.05} & \textbf{109} \\
    \cmidrule{1-14}
    
    \multirow{4.5}{*}{Test} & \multirow{2.5}{*}{Case} & Dementia & 23 & 23 & 0 & 78$\pm$7.4  & 11  & 12 & 4.15  & 20.89 & 21.86  & 19.50 & 21 \\ 
    \cmidrule{3-14}
     &  & MCI & 50 & 50 & 0 & 73.94$\pm$4.82  & 28  & 22 & 9.61  & 50.30 & 52.52  & 47.59 & 36 \\
    \cmidrule{2-14}
     & Control & HC & 77 & 77 & 0 & 68.68$\pm$2.91 & 36  & 41 & 14.68 & 94.93 & 100.31 & 94.13 & 1 \\
    
    \cmidrule{1-14}
    \textbf{SUBTOTAL} & --- & --- & \textbf{150} & \textbf{150} & \textbf{0} & \textbf{71.86$\pm$5.74} & \textbf{75} & \textbf{75} & \textbf{28.44} & \textbf{166.12} & \textbf{174.69} & \textbf{161.22} & \textbf{58} \\
    \cmidrule{1-14}
    
    \midrule
    
    \textbf{TOTAL} & --- & --- & \textbf{870} & \textbf{186} & \textbf{684} & \textbf{67.49$\pm$8.23} & \textbf{361} & \textbf{498} & \textbf{166.42} & \textbf{1066.33} & \textbf{1121.25} & \textbf{1069.27} & \textbf{167} \\
    
    \bottomrule
\end{tabular}
\end{table*}


\section{Neuropsychologically calibrated diagnostic label preparation}
\label{sec:neroCALlabels} 

\subsection{Analysis of neuropsychological performance of \cognospeak{}}

\begin{figure*}[!ht]
\centering
\subfigure[Comparison of \cognospeak{} neuropsychological scores with the FLAME dataset.]
{\label{fig:cogno_FLAME}
\includegraphics[width=\linewidth]{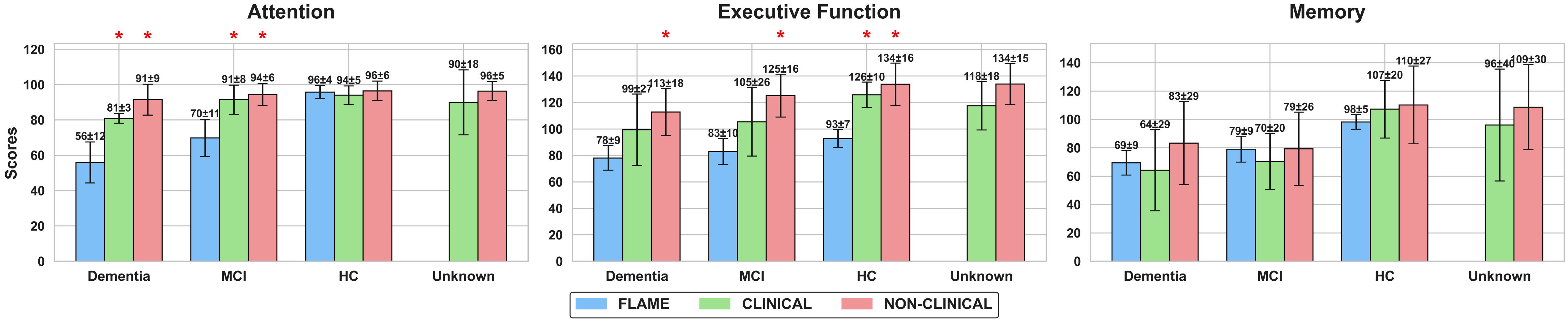}}
\subfigure[Distribution of neuropsychological performance across diagnostic groups.]
{\label{fig:cogno_hist_data_desc}
\includegraphics[width=\linewidth]{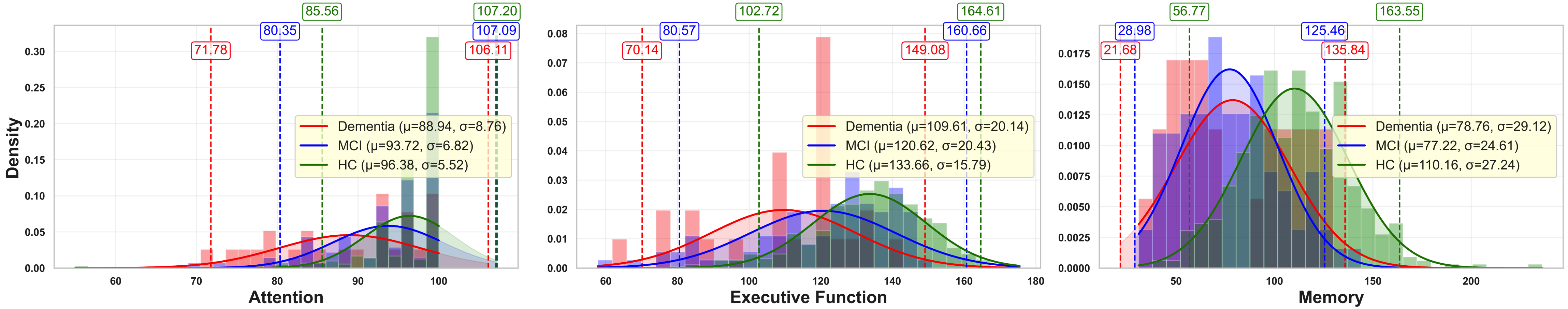}}
\subfigure[Comparison of neuropsychological scores derived from clinical and non-clinical sources.]
{\label{fig:cogno_raincloud_clinical}
\includegraphics[width=\linewidth]{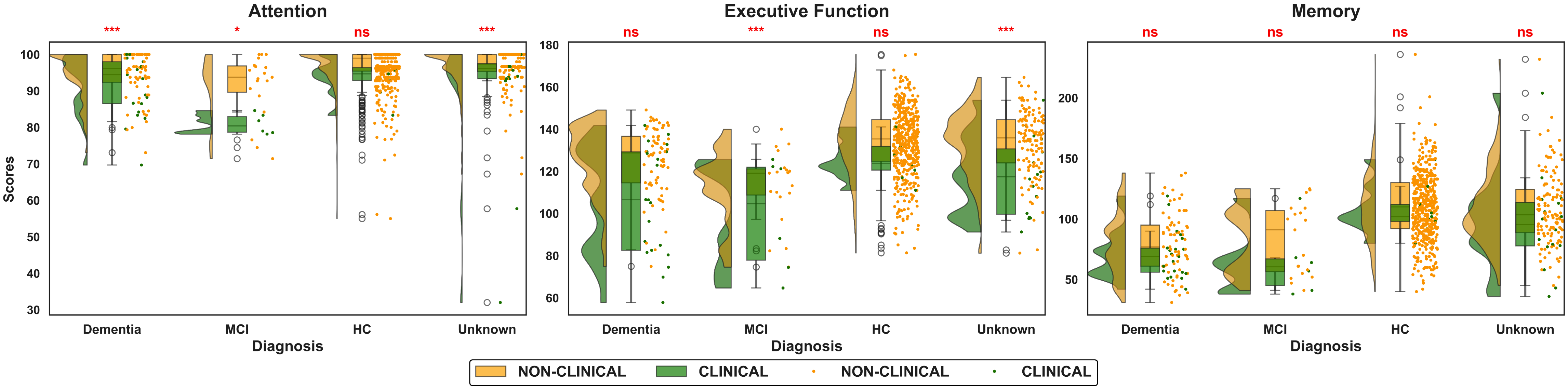}}
\caption{
\textbf{Domain-level neuropsychological performance analysis and dataset comparison.}
(a) Mean neuropsychological performance comparison between \cognospeak{} and the FLAME dataset. Bars indicate mean $\pm$ standard deviation. Red asterisks mark cases where between-source mean differences exceed the combined within-group variability, indicating pronounced distributional separation.
(b) Gaussian-fitted histograms illustrating score distributions for dementia, MCI, and healthy control (HC) participants across attention, executive function, and memory domains. Solid curves represent fitted normal distributions, while dashed lines denote 95\% confidence interval (CI) boundaries estimated as $\mu \pm 1.96\sigma$.
(c) Raincloud plots comparing neuropsychologically derived scores originating from clinically verified and non-clinical data sources. Statistical significance between sources within each diagnostic group was assessed using independent two-sample t-tests, with ns (not significant), $*$ ($p<0.1$), $**$ ($p<0.05$), and $***$ ($p<0.01$).
}
\label{fig:cognitive_performance_desc}
\vspace{-15pt}
\end{figure*}

Figure~\ref{fig:cognitive_performance_desc} provides a comprehensive statistical characterisation of neuropsychological performance distributions across three domains, datasets, and data sources.

\subsubsection{Against external \ FLAME dataset}

To assess the alignment between neuropsychological performance on the external FLAME study and on \cognospeak{}, we performed pairwise comparisons of mean neuropsychological scores across the attention, executive function and memory domains (Fig.~\ref{fig:cogno_FLAME}). 
For each diagnostic category (Dementia, MCI, and HC), FLAME cohorts were compared against both diagnostically certain (clinically verified) and uncertain (non-clinical) \cognospeak{} cohorts.
Because subject-level FLAME data were unavailable, formal inferential statistical testing could not be performed. Instead, we adopted a conservative descriptive separation criterion designed to highlight substantial distributional divergence. A comparison was marked with a red asterisk when the absolute difference between group means exceeded the combined within-group variability:
\begin{equation}
    |\mu_{\text{FLAME}} - \mu_{\text{target}}| > (\sigma_{\text{FLAME}} + \sigma_{\text{target}})
    \label{eq:cohort_separation}
\end{equation}

where $\text{target} \in \{\text{clinical}, \text{non-clinical}\}$. 
This condition identifies cases where shifts in central tendency surpass expected sampling variability, indicating large effect-size separation rather than inferential statistical significance.

Several attention and executive function conditions met this criterion, demonstrating systematic, domain-dependent scaling differences across datasets. In contrast, memory performance remained comparatively aligned across FLAME, and \cognospeak{}, suggesting that neuropsychologically anchored memory measures preserve consistent structural behaviour across independently collected cohorts and validating the distinctiveness of FLAME-generated labels against clinical benchmarks.

\subsubsection{Overall performance distributions within \cognospeak{}}

Overall neuropsychological performance distributions across dementia, MCI, and healthy control groups are illustrated using Gaussian-fitted histograms in Fig.~\ref{fig:cogno_hist_data_desc}. 
Solid curves represent fitted normal distributions, while dashed lines indicate estimated 95\% confidence intervals ($\mu \pm 1.96\sigma$), quantifying expected population variability.

Clear diagnostic separation is observed across all domains. 
Memory scores show strong differentiation between healthy participants and cognitively impaired groups, with mean values of 78.76 (Dementia), 77.22 (MCI), and 110.16 (HC). Although dementia and MCI exhibit similar mean performance, the dementia cohort displays a wider confidence interval, reflecting increased heterogeneity consistent with advanced cognitive decline.

Attention and executive function demonstrate progressively increasing performance across diagnostic stages. Attention accuracy means were 88.94\%, 93.72\%, and 96.38\% for Dementia, MCI, and HC, respectively, while executive function averages reached 109.61, 120.62, and 133.66. The displacement of distribution means across domains indicates a monotonic relationship between neuropsychological performance and disease severity.

Collectively, these distributions demonstrate that neuropsychologically derived scoring produces clinically meaningful separation between diagnostic groups, with memory impairment providing the strongest discriminative signal.

\subsubsection{Performance across clinical and non-clinical sources}

Potential bias arising from heterogeneous data provenance was evaluated by comparing clinically verified and non-clinical cohorts using raincloud visualisation (Fig.~\ref{fig:cogno_raincloud_clinical}). Independent two-sample $t$-tests were conducted within each diagnostic category across neuropsychology (attention, executive function and memory) domains to determine whether scores originated from the same underlying population distribution.
Missing values were handled using non-stratified listwise deletion to preserve the robustness of the resulting statistics.

Results indicate domain-dependent effects of the data source. Attention accuracy and executive function scores exhibit several statistically significant differences between clinical and non-clinical sources across Dementia, MCI, and Unknown groups. In contrast, memory scores remain largely non-significantly different across sources, indicating stability of memory-derived neuropsychological measures despite variation in acquisition context or supervision level. 
This robustness supports the use of memory-based neuropsychological thresholds for large-scale semi-automatic dataset curation and reinforces the suitability of neuropsychologically calibrated labelling under heterogeneous real-world data conditions.

\subsection{Label Reassignment Strategies}
\label{subsec:labelReassign}

Because diagnostic annotations originating from non-clinical recruitment sources were not independently clinically verified, they could not be assumed to represent reliable ground-truth supervision. Training directly on such labels risks propagating label noise into downstream models, a well-recognised limitation in real-world clinical machine learning \cite{frenay2013classification}.

To address this issue, we adopted a neuropsychologically calibrated label-reassignment framework in which diagnostic categories were inferred from objective neuropsychological-performance measures rather than relying solely on the original annotations. The resulting labels function as clinically grounded proxy annotations, conceptually aligned with weak-supervision approaches that improve supervisory quality using imperfect or partially reliable signals \cite{ratner2019snorkel}. Improved downstream performance when evaluated on clinically confirmed test labels was interpreted as evidence of closer approximation to latent clinical ground truth, consistent with prior work showing that label refinement improves generalisation under noisy-label conditions \cite{goldberger2017training}.

Rather than learning diagnostic boundaries purely from data, the proposed framework transfers small amounts of clinically validated knowledge into large-scale automatically labelled datasets through neuropsychologically anchored thresholding. Diagnostic cut-offs ($T_1$ and $T_2$) were derived using three complementary strategies: (1) clinically established thresholds from the FLAME cohort (\textit{FLAME}), (2) Gaussian midpoint calibration estimated from a clinically verified subset of \cognospeak{} (\textit{\CognoG{}}), and (3) normative deviation thresholds defined using z-score distributions relative to HC variability within \cognospeak{} (\textit{\CognoZ{}}). Together, these strategies balance clinical priors with data-driven estimation while preserving interpretability.

Two label-selection procedures were evaluated. In the first, labels were assigned solely from neuropsychological domain scores. In the second, Montreal Cognitive Assessment (MoCA) scores were incorporated as stabilising clinical priors when available. Participants with MoCA $\leq17$ were labelled as Dementia, scores between $18$--$25$ as MCI, and scores $>25$ as HC, consistent with established screening conventions.

\subsubsection{Strategy 1: FLAME Clinical Thresholding (\textit{FLAME})}

The first strategy transfers external clinical knowledge into large-scale datasets using reference statistics obtained from the FLAME clinical cohort. Mean neuropsychological scores for dementia ($\mu_D$), MCI ($\mu_M$), and HC ($\mu_H$) groups were extracted from each neuropsychological domain (memory, attention, and executive function). 

Diagnostic decision boundaries were defined as midpoints between
adjacent clinical group means by using Equation \ref{eq:thresholds}.

\begin{equation}
\begin{aligned}
T_1 &= \mu_D + \frac{\mu_M - \mu_D}{2}, \\
T_2 &= \mu_M + \frac{\mu_H - \mu_M}{2}.
\end{aligned}
\label{eq:thresholds}
\end{equation}

\subsubsection{Strategy 2: Cogno Gaussian Thresholding (\textit{\CognoG{}})}

Instead of relying on external clinical references, the second strategy
estimates diagnostic boundaries directly from a subset of trusted,
clinically validated participants.

Let $x_i^{(d)}$ denote participant $i$'s performance in neuropsychological
domain $d \in \{\text{memory},\text{attention},\text{executive}\}$.
Class-specific means ($\mu_D,\; \mu_M,\; \mu_H$) were estimated from trusted samples. 
Decision thresholds were defined as midpoints between adjacent
diagnostic distributions using Equation \ref{eq:thresholds}.

\subsubsection{Strategy 3: Cogno Z-score Thresholding (\textit{\CognoZ{}})}

A complementary strategy was developed using deviation from normative
healthy cognition. Let $\mu_{HC}$ and $\sigma_{HC}$ denote the mean and standard deviation of HC participants. Neuropsychological scores were standardised as

\begin{equation}
    z = \frac{x - \mu_{\text{HC}}}{\sigma_{\text{HC}}}
    \label{eq:z_score}
\end{equation}

Following common neuropsychological practice, impairment thresholds were
defined relative to population variability:

\begin{equation}
    T_{1} = \mu_{\text{HC}} - 2.5\sigma_{\text{HC}}, \qquad T_{2} = \mu_{\text{HC}} - 1.5\sigma_{\text{HC}}
    \label{eq:thresholds}
\end{equation}

Following established neuropsychological practice, impairment thresholds were defined relative to normative population variability rather than fixed diagnostic cut-offs. Scores with $z>-1.5$ were considered within normal limits, values between $-2.5<z\leq-1.5$ indicated mild cognitive impairment (MCI), and scores with $z\leq-2.5$ reflected substantial cognitive impairment. These thresholds were not intended to establish formal clinical diagnoses, but to provide clinically grounded priors for automated label generation. The selected boundaries are broadly consistent with neuropsychological criteria commonly used in MCI and dementia assessment, where performance approximately 1.5--2 standard deviations below normative expectations is considered clinically meaningful \cite{petersen1999mild, jak2009quantification}. Moderately conservative thresholds additionally improve class stability by avoiding the extreme rarity associated with severe deficit criteria (e.g., $z<-3$), thereby supporting more robust downstream model training and improved classification performance.

\subsubsection{Label reassignment process}

Following the calculation of threshold values ($T_1$ and $T_2$), participants were assigned diagnostic labels, $y_i$, based on their neuropsychological scores, $x_i$, according to Equation \ref{eq:label_reassgn}. 

\begin{equation}
y_i =
\begin{cases}
\text{Dementia}, & x_i \le T_1, \\
\text{MCI},      & T_1 < x_i \le T_2, \\
\text{HC},       & x_i > T_2.
\end{cases}
\label{eq:label_reassgn}
\end{equation}

This formulation assumes Gaussian distributions; FLAME thresholding uses population-level clinical statistics rather than dataset-specific data to approximate Bayesian decision boundaries. This reduces sensitivity to dataset heterogeneity while preserving clinical interpretability. Furthermore, using moderately conservative boundaries avoids extreme cut-offs (e.g., $z < -3$), which introduces meaningful priors that improve class balance and downstream model stability.

\begin{figure*}[!ht]
\centering
\includegraphics[width=\linewidth]{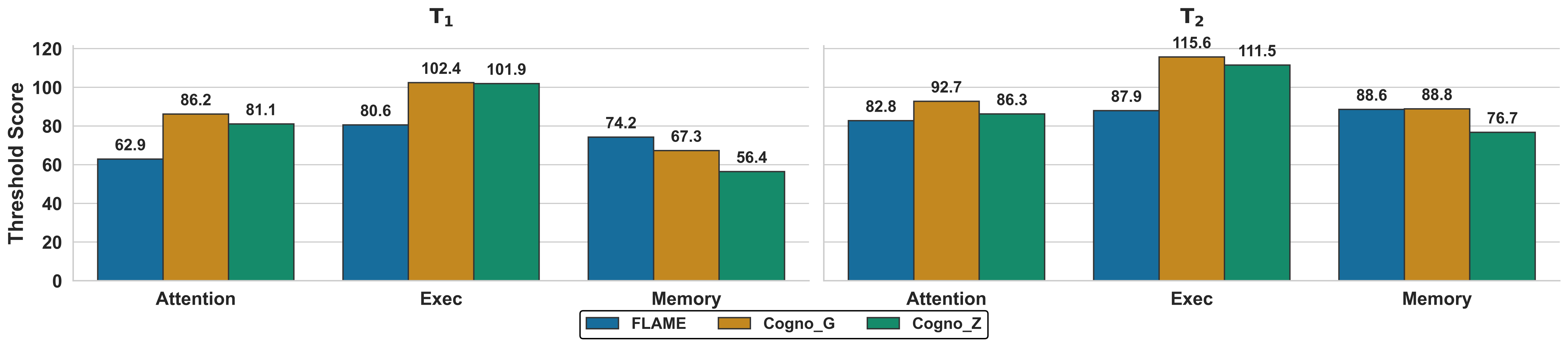}
\caption{\textbf{Diagnostic thresholds by three sorting strategies:} Clinically anchored diagnostic thresholds derived using FLAME clinical references, data-calibrated Gaussian midpoints, and healthy-control z-score deviations across memory, attention, and executive function domains. The figure illustrates how small amounts of trusted clinical information may inform decision boundaries.}
\label{fig:thresh_diag}
\end{figure*}

Figure \ref{fig:thresh_diag} shows the estimated $T_1$ and $T_2$ threshold values. 
Across neuropsychological domains, these approaches produced distinct decision boundaries. For example, memory-domain dementia thresholds ranged from 67.30 (\CognoG{}) to 76.75
(\CognoZ{}), while attention-domain thresholds shifted toward higher performance ranges (86.18--86.25), reflecting domain-specific separability. Executive function thresholds showed the largest difference (80.58--111.46), indicating strong scaling
differences between externally derived clinical priors and dataset-adapted estimates.
These results demonstrate that incorporating clinically informed
decision boundaries substantially reduce label noise and enable
weakly supervised models to achieve improved diagnostic
classification performance.

\begin{figure*}[!ht]
\centering
\includegraphics[width=\linewidth]{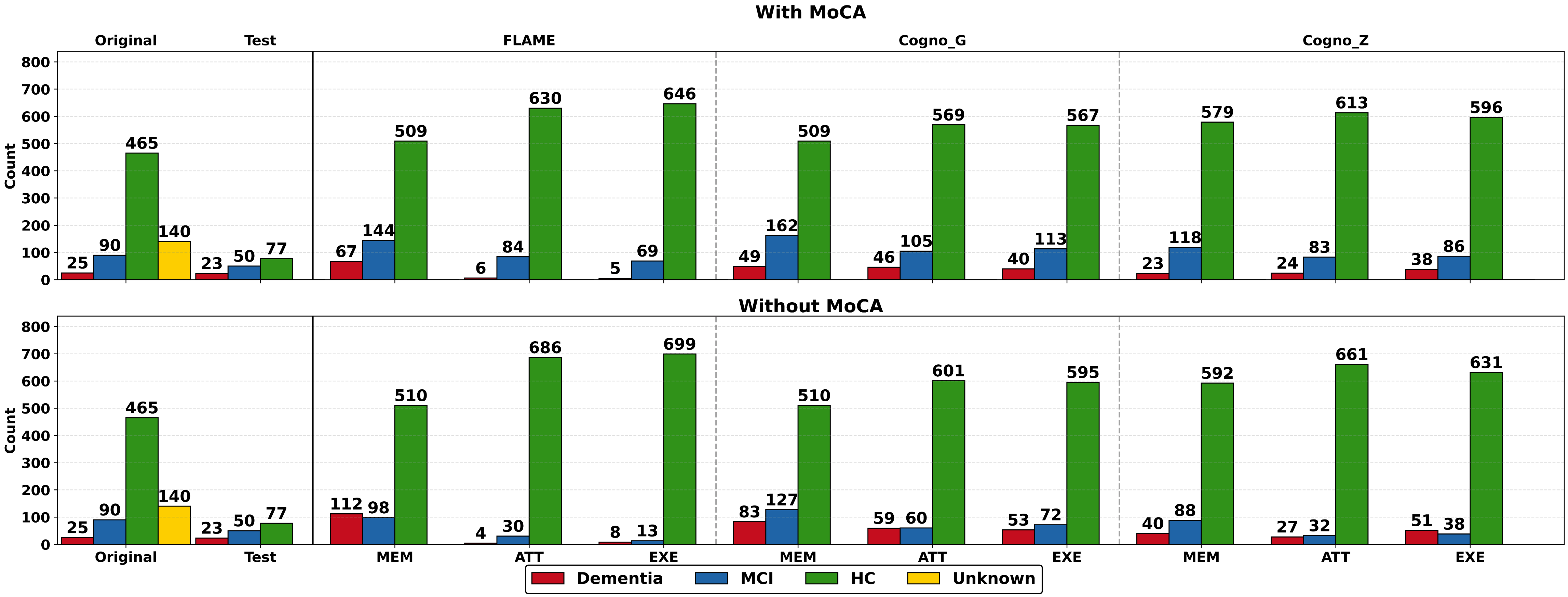}
\caption{
\textbf{Distribution of diagnostic labels generated using neuropsychologically calibrated reassignment strategies.}
The figure compares class distributions obtained from the original dataset annotations, the internal \cognospeak{} test cohort, and automatically reassigned diagnostic labels produced using three thresholding strategies (FLAME clinical reference thresholds, Cogno Gaussian midpoint thresholds, and Cogno Z-score normative deviation thresholds) across memory (MEM), attention (ATT), and executive function (EXE) domains.
The top row shows label distributions when Montreal Cognitive Assessment (MoCA) scores were incorporated as an auxiliary clinical override, whereas the bottom row shows label assignments derived purely from cognition, without MoCA information. Bars represent counts of participants assigned to Dementia, Mild Cognitive Impairment (MCI), Healthy Control (HC), and Unknown categories.
Across strategies, neuropsychologically anchored reassignment reduces reliance on uncertain original annotations while preserving clinically plausible class proportions. The comparison highlights how different thresholding assumptions influence class balance and demonstrates the stabilising effect of incorporating clinical priors during large-scale automatic label construction.
}
\label{fig:labels_dist}
\vspace{-10pt}
\end{figure*}

Figure~\ref{fig:labels_dist} summarises the distributional effects of the proposed label reassignment strategies. The original annotations exhibit a strong imbalance toward the MCI category, whereas all strategies systematically redistribute participants toward healthier classifications, producing class proportions more consistent with the external test cohort. Across methods, memory-based reassignment yields the most stable distributions, while attention- and executive-based thresholds introduce larger variability depending on the underlying calibration strategy. Incorporating MoCA priors consistently stabilises class allocation, reducing extreme shifts in diagnostic proportions and improving alignment across domains. Notably, Z-score thresholding produces the most conservative impairment estimates, assigning fewer dementia labels and expanding the healthy-control group, whereas Gaussian midpoint approaches preserve intermediate MCI representation. Overall, these patterns indicate that clinically anchored relabeling primarily acts as a label-denoising mechanism, correcting potential over-assignment of impairment present in the original dataset while maintaining coherent diagnostic structure across neuropsychological domains.




\subsection{Experimental Setup}

Our previous research indicated superior performance from transformer-based models; thus transformer-based models such as \bart{} \cite{lewis2019bart}, \roberta{} \cite{liu2019roberta} and \distilbert{} \cite{sanh2019distilbert} are the only classifiers used in this study \cite{pahar2025cognospeak,
pahar2025mutlimodalfusion, pahar2026can}. 
We have adopted a similar transfer-learning approach to our recent research while using the original labels for training \cite{pahar2025mutlimodalfusion}. 
The transformer-based models were further fine-tuned on the newly assigned diagnostic labels from three strategies, and tested on the held-out test set comprised of 150 participants, as shown in B part of Figure \ref{fig:summary_paper} and Table \ref{table:dataset}. 
As the data is imbalanced (Table \ref{table:dataset}), macro $F_1$-score is used as the primary evaluation metric. 
A majority voting algorithm was used to predict the final label for the question combinations.

\section{Results}
\label{sec:results}

\subsection{Binary classification}

\begin{figure*}[!ht]
\centering
\subfigure[Task-level $F_1$-score heatmap across labeling strategies.]
{\label{fig:2way_heatmap}
\includegraphics[width=\linewidth]{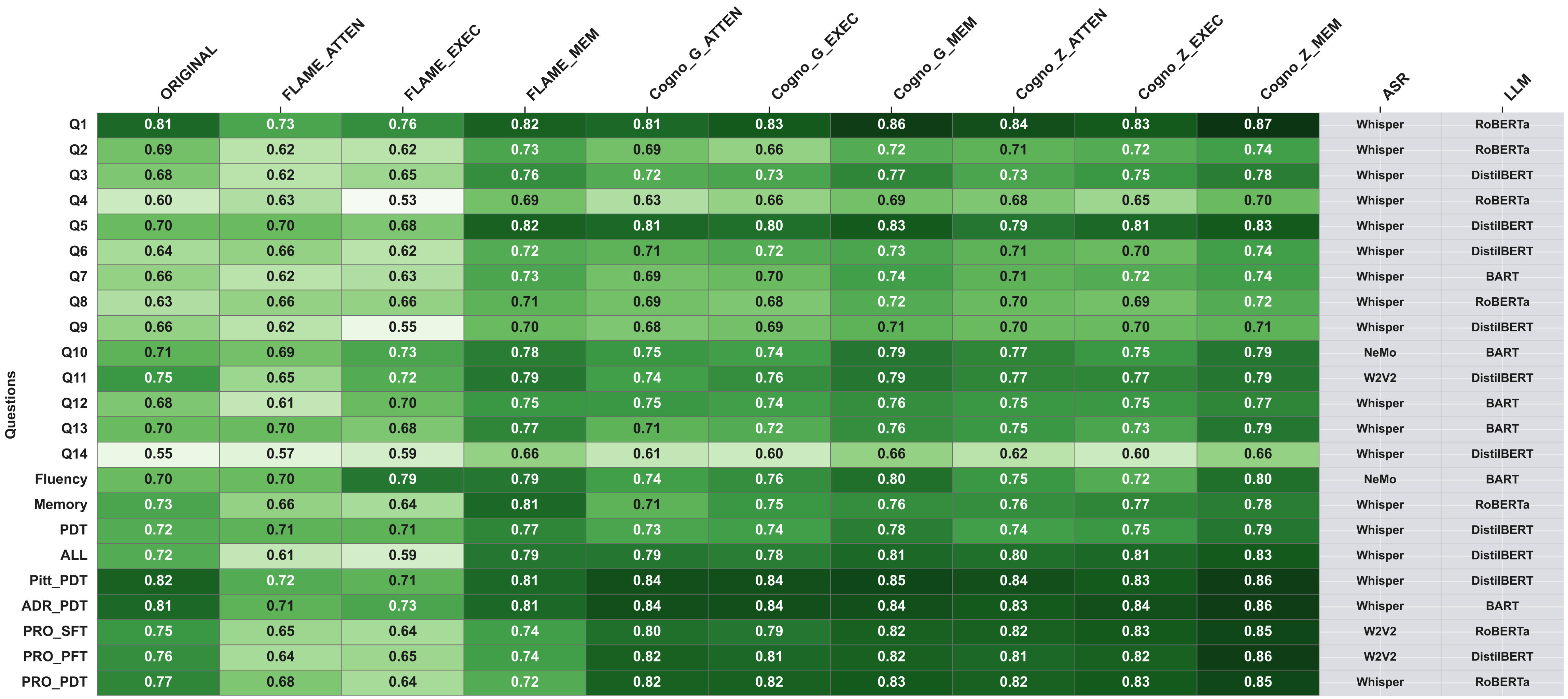}}
\subfigure[Effect of MoCA-based clinical priors on classification performance.]
{\label{fig:2way_moca}
\includegraphics[width=\linewidth]{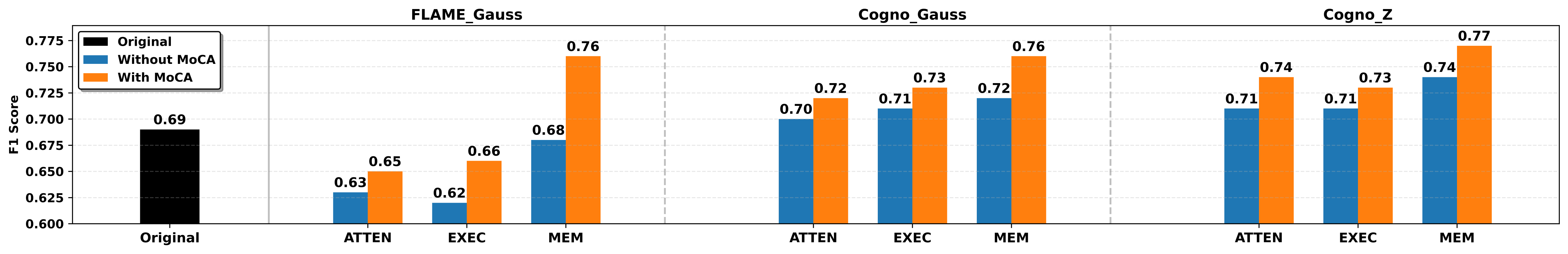}}
\subfigure[Distribution of $F_1$-scores across labeling strategies.]
{\label{fig:2way_boxplot}
\includegraphics[width=0.48\linewidth]{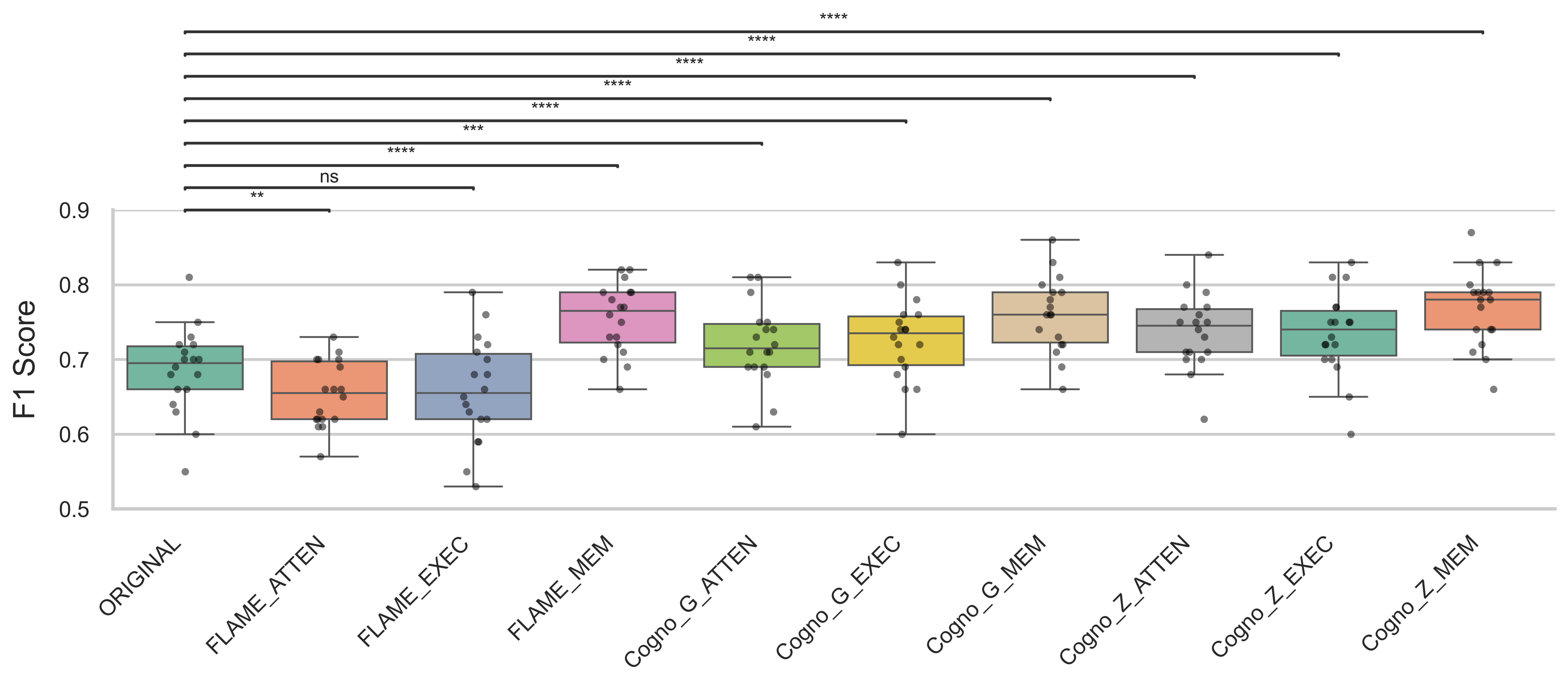}}
\subfigure[Confusion matrix using original diagnostic labels.]
{\label{fig:2way_conf_ori}
\includegraphics[width=0.24\linewidth]{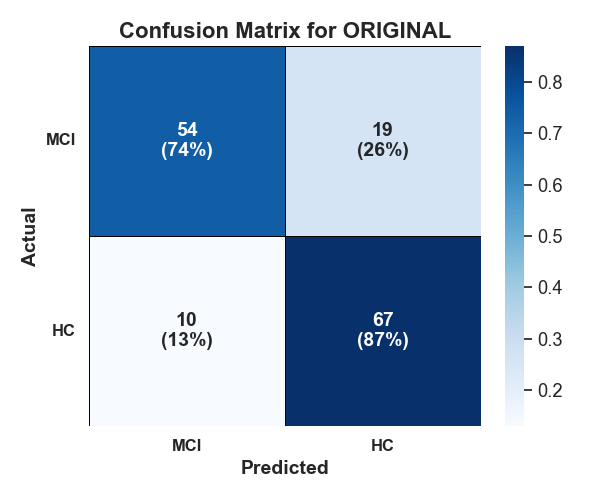}}
\subfigure[Confusion matrix using Cogno-Z memory-derived labels.]
{\label{fig:2way_conf_cogno}
\includegraphics[width=0.24\linewidth]{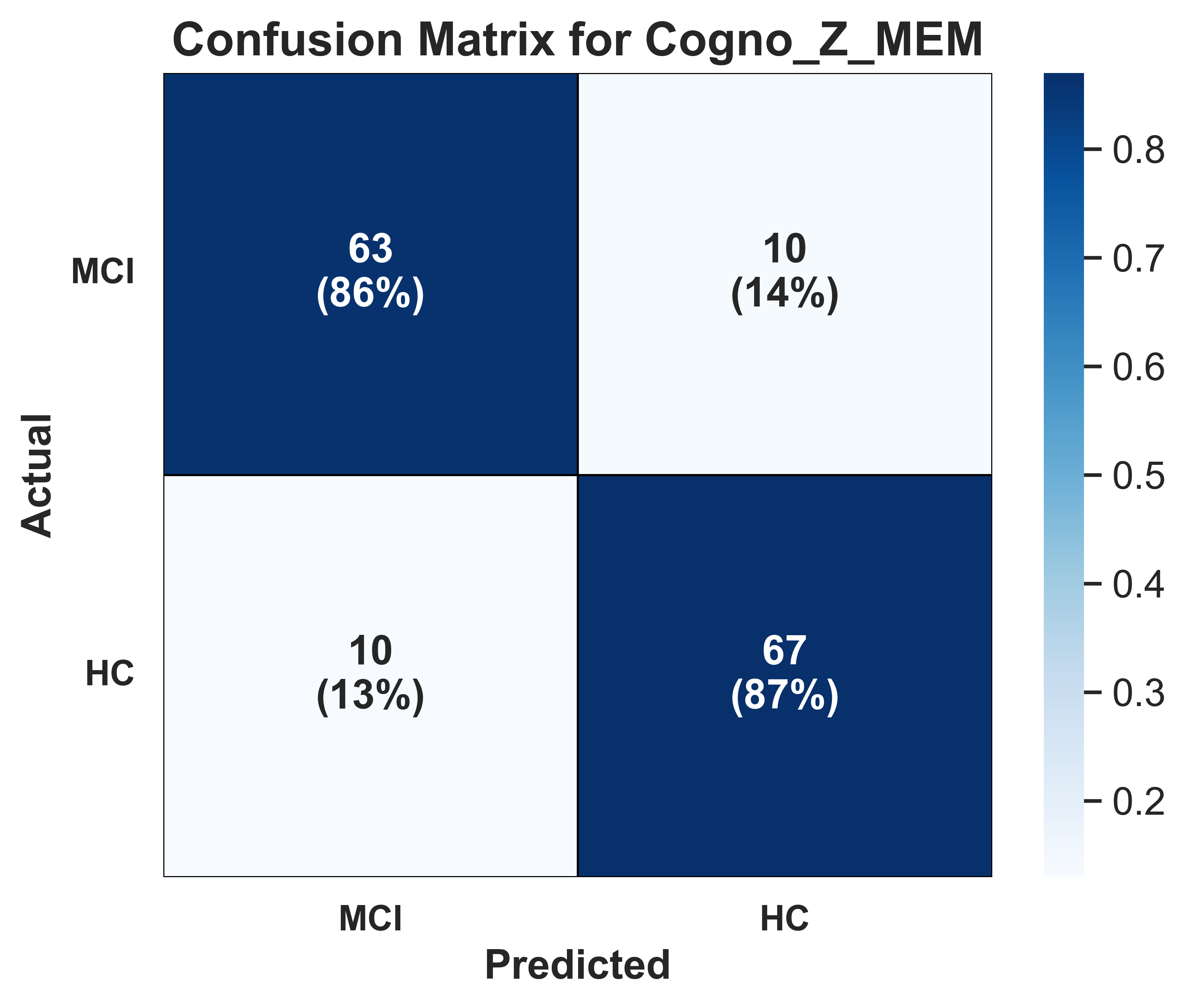}}
\caption{
\textbf{Summary of binary cognitive decline classification performance.}
(a) Heatmap of $F_1$-scores across individual speech tasks comparing original clinical labels with proposed neuropsychologically informed labelling strategies.
Columns represent FLAME clinical thresholds, data-driven Gaussian calibration (Cogno-G), and normative z-score calibration (Cogno-Z).
Associated automatic speech recognition (ASR) and transformer-based model configurations are shown for reference.
(b) Average classification performance obtained with and without incorporation of MoCA-derived clinical priors.
MoCA integration produces consistent performance gains across domains, supporting the role of global cognitive screening as a stabilising supervisory signal.
(c) Distribution of $F_1$-scores across labelling strategies, demonstrating a systematic upward shift in performance relative to the original labels.
(d) Confusion matrix obtained using original diagnostic labels, showing increased misclassification between impaired and healthy participants.
(e) Confusion matrix obtained using Cogno-Z memory-based labels, illustrating improved class separation and reduced diagnostic ambiguity.
}
\label{fig:2-way-class-summary}
\end{figure*}

Across binary classification experiments, neuropsychologically informed label-generation strategies consistently outperformed the original diagnostic annotations (Fig.~\ref{fig:2way_heatmap}). 
The highest performance was achieved using the \CognoZ{} memory-based (\CognoZ{}\_MEM) labelling strategy, reaching an $F_1$-score of 0.87 for the $Q_1$ task compared with 0.81 obtained using the original labels.

Performance gains were particularly pronounced for cognitively demanding speech tasks. The largest improvements were observed for $Q_5$ (0.70 $\rightarrow$ 0.83), and the combination of all questions (0.72 $\rightarrow$ 0.83), i.e. +13\% and +11\%, suggesting that structured memory elicitation tasks provide strong signals of cognitive impairment.
Conversely, $Q_{14}$ remained challenging across all methods, indicating limited diagnostic information content for this task. 

Across labelling approaches, memory-derived thresholds consistently outperformed attention and executive function variants, supporting the hypothesis that memory impairment represents the most robust speech-derived marker of cognitive decline. 

Analysis across automatic speech recognition systems showed that Whisper-based transcriptions produced the most stable performance across tasks, while NeMo and W2V2 results exhibited greater variability. Similarly, RoBERTa classifiers generally achieved higher peak performance compared with DistilBERT and BART, although improvements were secondary to the effect of label refinement itself.

Distributional analysis (Fig.~\ref{fig:2way_boxplot}) demonstrates a global upward shift in $F_1$-score distributions under neuropsychologically calibrated labelling, indicating that gains were systematic rather than driven by isolated experiments.
Confusion-matrix comparisons (Figs.~\ref{fig:2way_conf_ori} and \ref{fig:2way_conf_cogno}) further reveal reduced misclassification and improved sensitivity to impaired participants.

\subsection{Multiclass classification}

\begin{figure*}[!ht]
\centering

\subfigure[Task-level multiclass classification performance.]
{\label{fig:3way_heatmap}
\includegraphics[width=\linewidth]{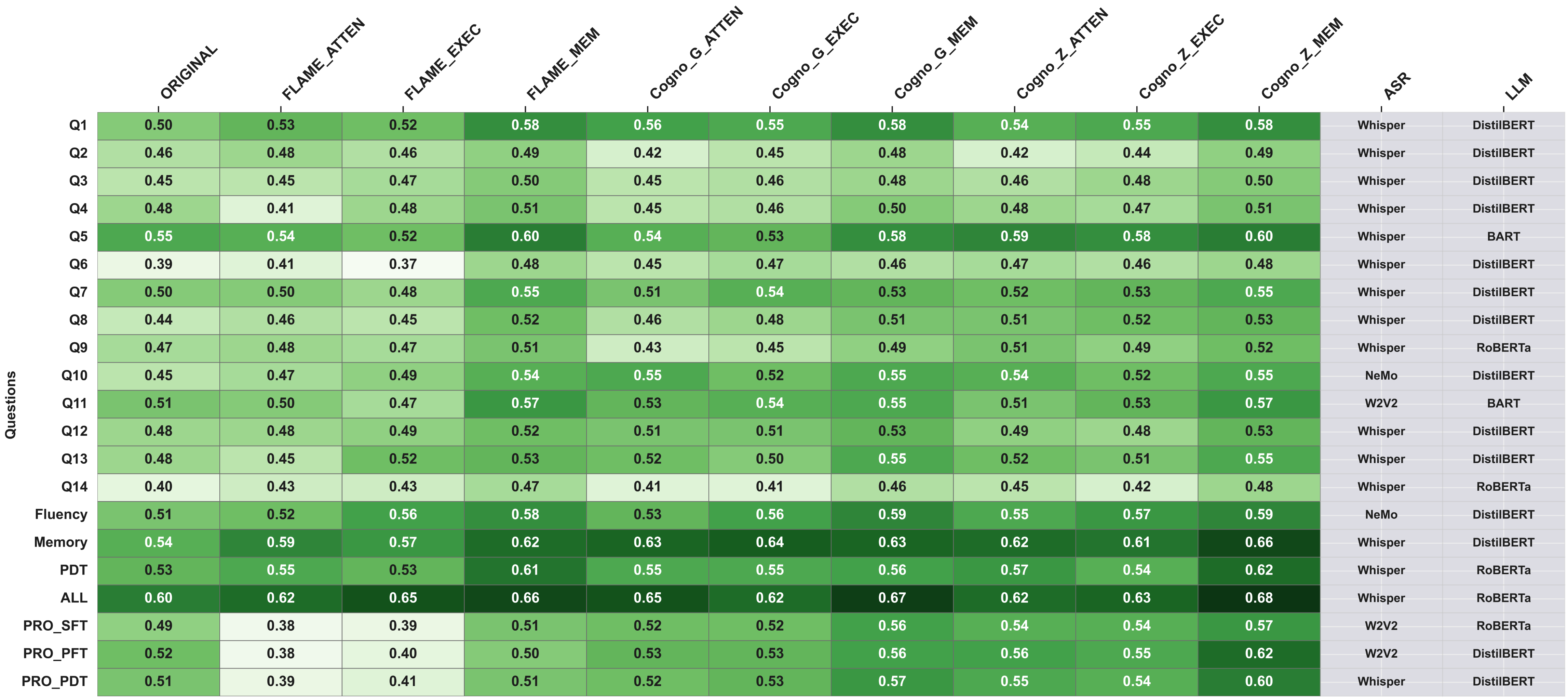}}
\subfigure[Impact of MoCA integration on multiclass classification.]
{\label{fig:3way_moca}
\includegraphics[width=\linewidth]{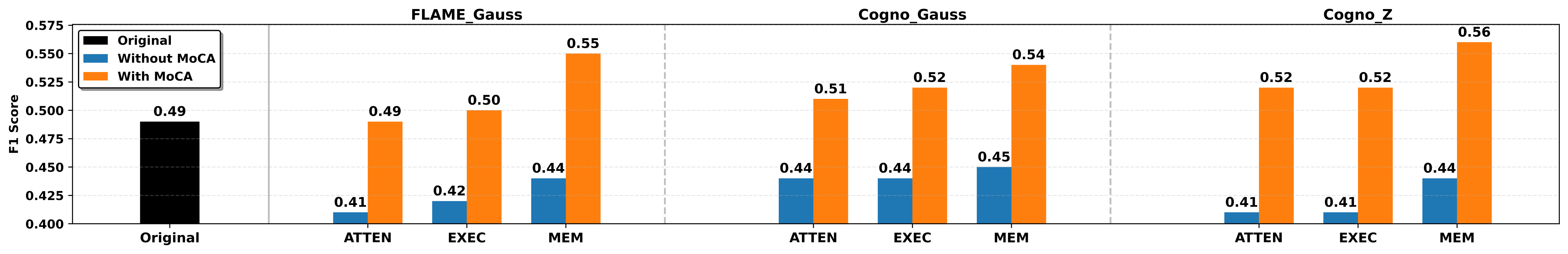}}
\subfigure[Distribution of $F_1$-scores across labeling strategies.]
{\label{fig:3way_boxplot}
\includegraphics[width=0.48\linewidth]{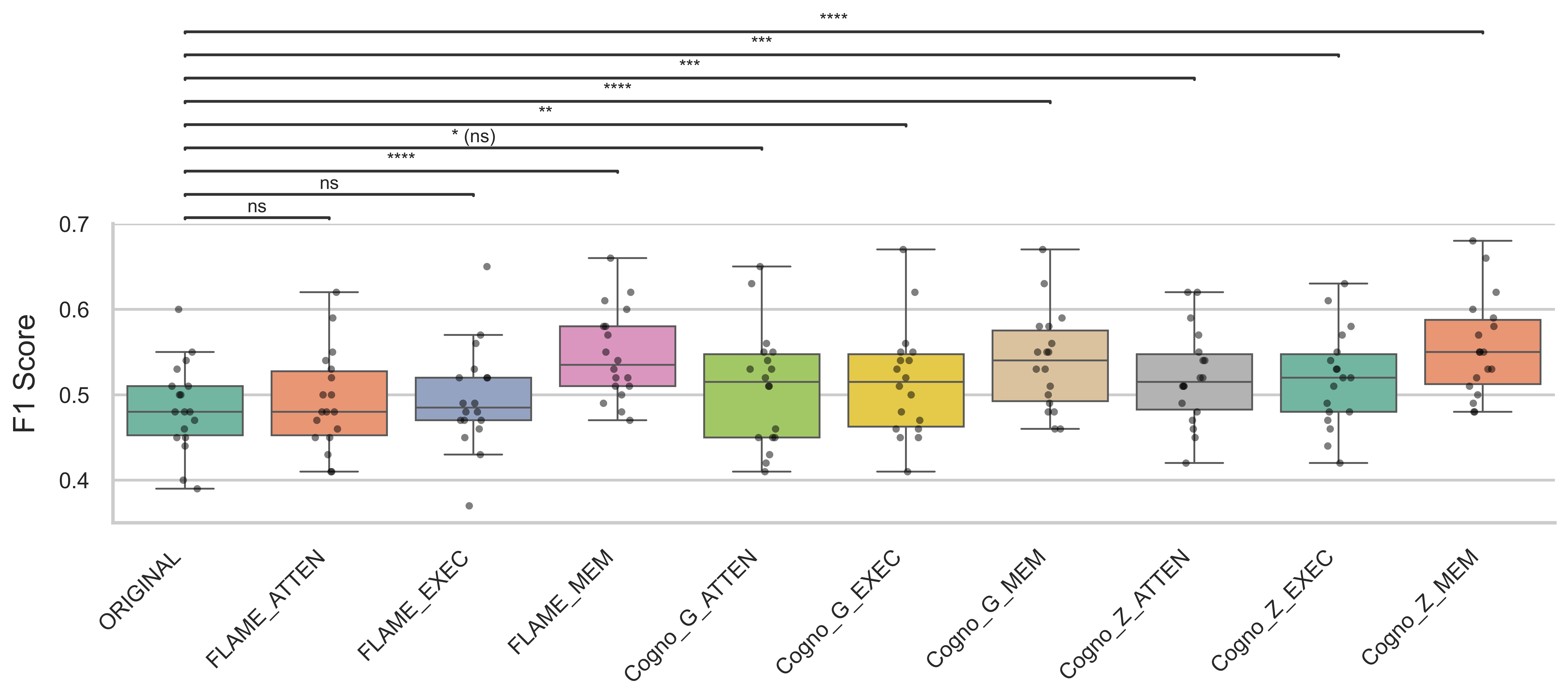}}
\subfigure[Confusion matrix using original diagnostic labels.]
{\label{fig:3way_conf_ori}
\includegraphics[width=0.24\linewidth]{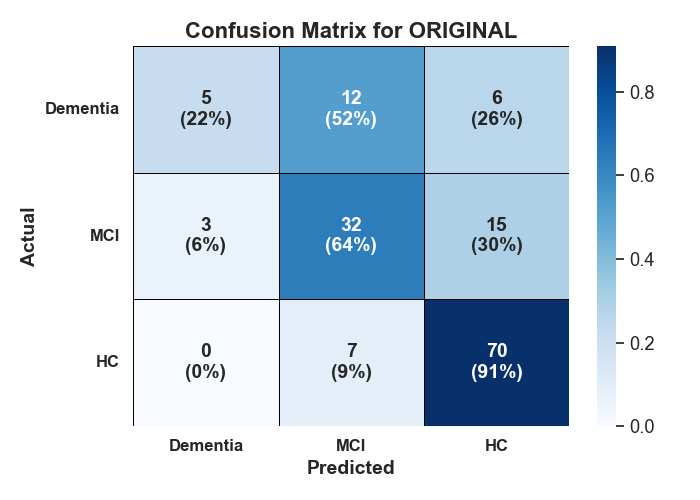}}
\subfigure[Confusion matrix using Cogno-Z memory-derived labels.]
{\label{fig:3way_conf_cogno}
\includegraphics[width=0.24\linewidth]{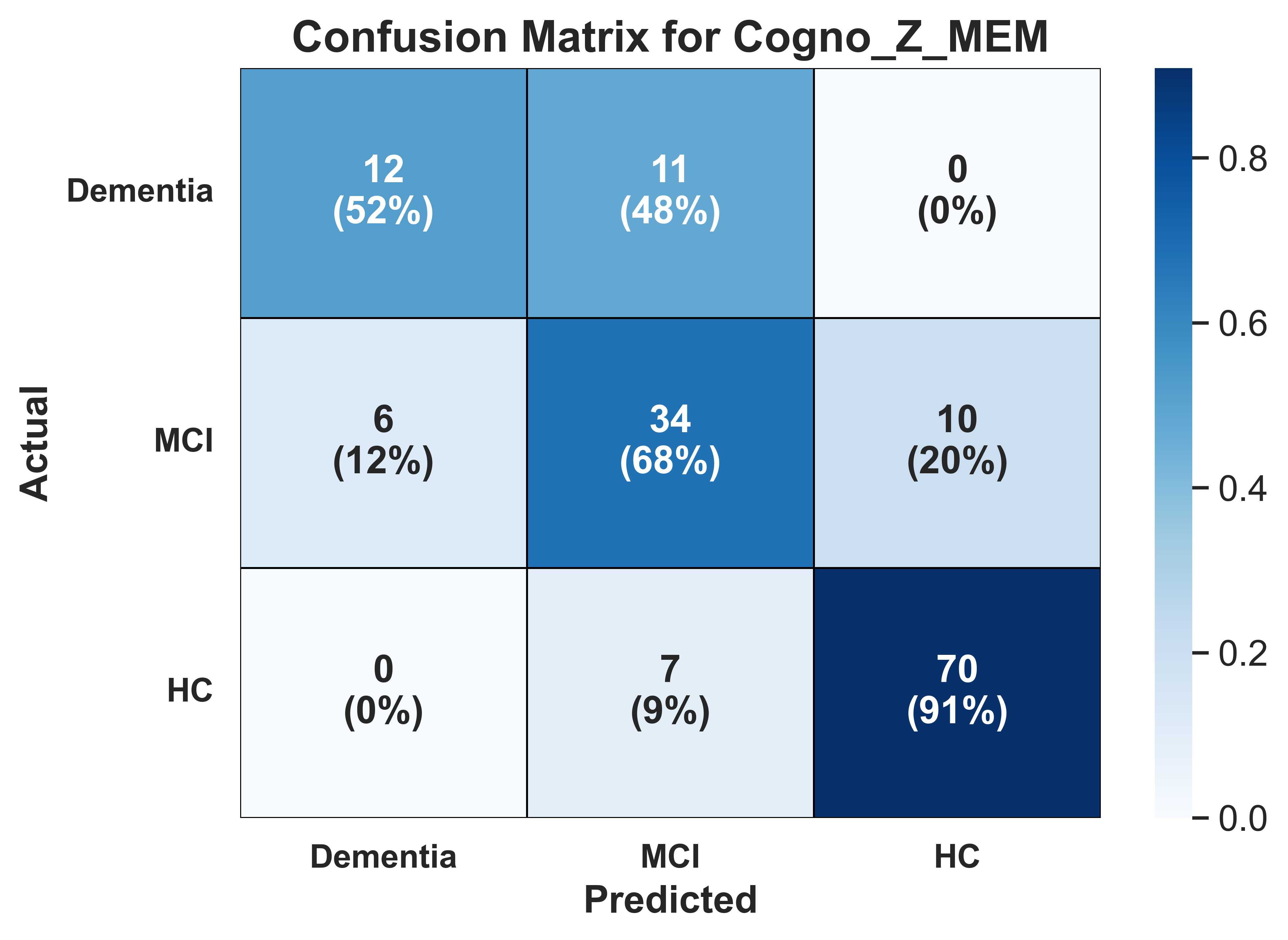}}
\caption{
\textbf{Summary of multiclass cognitive decline classification performance.}
(a) Heatmap showing multiclass $F_1$-scores across speech tasks and labelling strategies.
Neuropsychologically calibrated labels consistently outperform original annotations, particularly for memory-intensive tasks.
(b) Comparison of performance with and without MoCA-derived priors, demonstrating improved stability when separating dementia, MCI, and healthy control groups.
(c) Distributional comparison of $F_1$-scores illustrating global performance improvements achieved through clinically anchored thresholding.
(d) Confusion matrix obtained using original labels, highlighting difficulty distinguishing intermediate cognitive states.
(e) Confusion matrix obtained using Cogno-Z memory-derived labels, showing improved discrimination between dementia, MCI, and healthy controls.
}
\label{fig:3-way-class-summary}
\end{figure*}

Three-class classification constitutes a substantially more challenging setting due to the inclusion of the intermediate MCI category. Nevertheless, neuropsychologically guided labelling again improved performance relative to the original diagnostic labels (Fig.~\ref{fig:3way_heatmap}).

Using the \CognoZ\_MEM strategy, the highest $F_1$-score have been achieved using the `ALL' task, with a performance increase of +8\%, 0.60 to 0.68. 
The largest task-specific gains were observed for the Memory task (0.54 $\rightarrow$ 0.66) and the PDT task (0.53 $\rightarrow$ 0.62), indicating improved separation between dementia and MCI participants.

While improvements were smaller than in the binary setting, they remained consistent across nearly all tasks, suggesting that clinically anchored thresholding produces more reliable multiclass diagnostic boundaries.

ASR and language-model effects were again secondary to labeling strategy. Whisper transcriptions yielded the most stable performance, while RoBERTa and DistilBERT classifiers showed comparable behaviour across tasks. These findings indicate that performance gains primarily arise from improved training labels rather than architectural differences.

Confusion-matrix analysis (Figs.~\ref{fig:3way_conf_ori} and \ref{fig:3way_conf_cogno}) demonstrates improved discrimination between dementia and MCI classes, supporting the clinical validity of neuropsychologically informed labelling.

\subsection{Improvement due to MoCA}

Incorporating MoCA-derived clinical priors produced consistent performance improvements across all labelling strategies (Figs.~\ref{fig:2-way-class-summary}b and
\ref{fig:3-way-class-summary}b).
Paired statistical testing demonstrated that models trained with MoCA-informed labels achieved significantly higher $F_1$-scores than models trained without MoCA information
($t=15.65$, $p=2.77\times10^{-7}$, paired t-test).

Across tasks, MoCA integration increased performance by an average $\Delta$$F_1$ of 0.094, corresponding to an approximate 9--10\% relative improvement.
The effect size was extremely large (Cohen’s $d=5.22$), indicating that the improvement is not task-specific but systematic across experimental conditions.

These results suggest that global cognitive screening scores act as clinically meaningful priors that stabilise automatically generated diagnostic labels, particularly when distinguishing intermediate cognitive states such as mild cognitive impairment
(MCI).

Due to the lack of MoCA scores for all diagnostic classes, using MoCA alone to relabel the diagnosis has not been adopted. 
The selection strategy `without MoCA' was only adapted to internal validation, as it didn't produce better results.

\begin{table*}[!ht]
\centering
\footnotesize
\setlength{\tabcolsep}{6pt}
\caption{Performance Comparison across Internal and External Validation Cohorts (Binary vs. Multiclass Classification). The strategies and the neuropsychological domains performing significantly better are shown in bold.}
\label{tab:combined_results}
\begin{tabular}{llllcccc}
\toprule
Validation Type & Dataset & Size ($N$) & Task & Method & Mean\_diff ($\Delta F_1$) & Cohens\_d & Result \\
\midrule
\multirow{20}{*}{\textbf{Internal}} & \multirow{19.5}{*}{\cognospeak{}} & \multirow{9}{*}{\begin{tabular}{c} Case: 73 \\ Control: 77 \end{tabular}} & \multirow{9}{*}{Binary} & FLAME\_ATTEN    & -0.032 & -0.697 & 0.008 (**) \\
& \multirow{20}{*}{(Table \ref{table:dataset})} & & & FLAME\_EXEC              & -0.027 & -0.482 & 0.057 (ns) \\
& & & & \textbf{FLAME\_MEM}      & 0.070  & 2.635  & $<$0.001 (****) \\
\cmidrule{5-8}
& & & & \textbf{Cogno\_G\_ATTEN} & 0.035  & 1.029  & $<$0.001 (***) \\
& & & & \textbf{Cogno\_G\_EXEC}  & 0.041  & 1.379  & $<$0.001 (****) \\
& & & & \textbf{Cogno\_G\_MEM}   & 0.075  & 2.707  & $<$0.001 (****) \\
\cmidrule{5-8}
& & & & \textbf{Cogno\_Z\_ATTEN} & 0.053  & 2.352  & $<$0.001 (****) \\
& & & & \textbf{Cogno\_Z\_EXEC}  & 0.049  & 1.990  & $<$0.001 (****) \\
& & & & \textbf{Cogno\_Z\_MEM}   & \textbf{0.083}  & 3.306  & $<$0.001 (****) \\
\cmidrule{3-8} \addlinespace
& & \multirow{9}{*}{\begin{tabular}{c} Dementia: 23 \\ MCI: 50 \\ HC: 77 \end{tabular}} & \multirow{9}{*}{Multiclass} & FLAME\_ATTEN             & 0.007  & 0.272  & 0.264 (ns) \\
& & & & FLAME\_EXEC              & 0.011  & 0.393  & 0.228 (ns) \\
& & & & \textbf{FLAME\_MEM}      & 0.061  & 3.107  & $<$0.001 (****) \\
\cmidrule{5-8}
& & & & Cogno\_G\_ATTEN          & 0.023  & 0.572  & 0.080 (ns) \\
& & & & \textbf{Cogno\_G\_EXEC}  & 0.029  & 0.763  & 0.009 (**) \\
& & & & \textbf{Cogno\_G\_MEM}   & 0.053  & 2.024  & $<$0.001 (****) \\
\cmidrule{5-8}
& & & & \textbf{Cogno\_Z\_ATTEN} & 0.035  & 1.062  & $<$0.001 (***) \\
& & & & \textbf{Cogno\_Z\_EXEC}  & 0.033  & 1.136  & $<$0.001 (***) \\
& & & & \textbf{Cogno\_Z\_MEM}   & \textbf{0.069}  & 2.822  & $<$0.001 (****) \\
\midrule \addlinespace
\multirow{43}{*}{\textbf{External}} & \multirow{9}{*}{Pitt \cite{becker1994natural}} & \multirow{9}{*}{\begin{tabular}{c} Case: 194 \\ Control: 98 \end{tabular}} & \multirow{9}{*}{Binary} & FLAME\_ATTEN    & -0.100 & ---    & --- \\
& & & & FLAME\_EXEC     & -0.110 & ---    & --- \\
& & & & FLAME\_MEM     & -0.010 & ---    & --- \\
\cmidrule{5-8}
& & & & Cogno\_G\_ATTEN & 0.020  & ---    & --- \\
& & & & Cogno\_G\_EXEC  & 0.020  & ---    & --- \\
& & & & Cogno\_G\_MEM   & 0.030  & ---    & --- \\
\cmidrule{5-8}
& & & & Cogno\_Z\_ATTEN & 0.020  & ---    & --- \\
& & & & Cogno\_Z\_EXEC  & 0.010  & ---    & --- \\
& & & & \textbf{Cogno\_Z\_MEM}   & \textbf{0.040}  & ---    & --- \\
\cmidrule{2-8} \addlinespace
& \multirow{9}{*}{\adress{} \cite{luz2020alzheimer}} & \multirow{9}{*}{\begin{tabular}{c} Case: 24 \\ Control: 24 \end{tabular}} & \multirow{9}{*}{Binary} & FLAME\_ATTEN    & -0.100 & ---    & --- \\
& & & & FLAME\_EXEC     & -0.080 & ---    & --- \\
& & & & FLAME\_MEM    & 0.000  & ---    & --- \\
\cmidrule{5-8}
& & & & Cogno\_G\_ATTEN & 0.030  & ---    & --- \\
& & & & Cogno\_G\_EXEC  & 0.030  & ---    & --- \\
& & & & Cogno\_G\_MEM  & 0.030  & ---    & --- \\
\cmidrule{5-8}
& & & & Cogno\_Z\_ATTEN & 0.020  & ---    & --- \\
& & & & Cogno\_Z\_EXEC  & 0.030  & ---    & --- \\
& & & & \textbf{Cogno\_Z\_MEM}   & \textbf{0.050}  & ---    & --- \\
\cmidrule{2-8} \addlinespace
& \multirow{20}{*}{PROCESS-2 \cite{pahar2026PROCESS2}} & \multirow{9}{*}{\begin{tabular}{c} Case: 200 \\ Control: 200 \end{tabular}} & \multirow{9}{*}{Binary} & FLAME\_ATTEN    & -0.103 & --- & --- \\
& & & & FLAME\_EXEC     & -0.117 & --- & --- \\
& & & & FLAME\_MEM               & -0.027 & --- & --- \\
\cmidrule{5-8}
& & & & Cogno\_G\_ATTEN & 0.053  & ---  & --- \\
& & & & Cogno\_G\_EXEC  & 0.047  & ---  & --- \\
& & & & Cogno\_G\_MEM  & 0.063  & --- & --- \\
\cmidrule{5-8}
& & & & Cogno\_Z\_ATTEN & 0.057  & ---  & --- \\
& & & & Cogno\_Z\_EXEC & 0.067  & ---  & --- \\
& & & & \textbf{Cogno\_Z\_MEM}   & \textbf{0.093}  & ---  & --- \\
\cmidrule{3-8} \addlinespace
& & \multirow{9}{*}{\begin{tabular}{c} Dementia: 50 \\ MCI: 150 \\ HC: 200 \end{tabular}} & \multirow{9}{*}{Multiclass} & FLAME\_ATTEN    & -0.123 & --- & --- \\
& & & & FLAME\_EXEC     & -0.107 & --- & --- \\
& & & & FLAME\_MEM               & 0.000  & ---  & --- \\
\cmidrule{5-8}
& & & & Cogno\_G\_ATTEN          & 0.017  & ---  & --- \\
& & & & Cogno\_G\_EXEC           & 0.020  & ---  & --- \\
& & & & Cogno\_G\_MEM            & 0.057  & ---  & --- \\
\cmidrule{5-8}
& & & & Cogno\_Z\_ATTEN & 0.043  & ---  & --- \\
& & & & Cogno\_Z\_EXEC           & 0.037  & ---  & --- \\
& & & & \textbf{Cogno\_Z\_MEM}   & \textbf{0.090}  & ---  & --- \\
\bottomrule
\end{tabular}
\end{table*}

\subsection{External validation}

To evaluate the generalisability of the proposed neuropsychological relabelling strategies, models trained using reassigned labels were additionally evaluated on three independent external datasets: the Pitt Corpus \cite{becker1994natural}, \adress{} \cite{luz2020alzheimer}, and PROCESS-2 \cite{pahar2026PROCESS2}. Both Pitt and \adress{} corpus differ substantially from \cognospeak{} in participant demographics, recording conditions, task design, and transcription pipelines, providing a rigorous assessment of cross-cohort robustness.
They contained only two classes (cases and controls), thus they have been used only for binary classification (Figure \ref{fig:2way_heatmap}). As these two cohorts contained only picture description tasks (PDT), only $Q_{12}$ of \cognospeak{} was used for training (acronymed as Pitt\_PDT and ADR\_PDT). 

PROCESS-2 is compiled using the \cognospeak{} platform, and there was no subject overlap between 720 train set (table \ref{table:dataset}) and 400 test set from PROCESS-2. 
Only $Q_{10}$, $Q_{11}$, and $Q_{12}$ of \cognospeak{} was used for training. 
This diagnostic-rich data contains three classes, thus it has been used for both binary and multiclass classification in Figure \ref{fig:2way_heatmap} and \ref{fig:3way_heatmap} (acronymed as PRO\_SFT, PRO\_PFT and PRO\_PDT).

Across all external cohorts, memory-based relabelling strategies consistently achieved the strongest performance. On the Pitt dataset, the best-performing model was \textbf{Cogno\_Z\_MEM}, improving binary classification performance from $F_1=0.82$ to $0.86$ ($\Delta F_1=0.04$), while FLAME-based attention and executive-function relabelling substantially reduced performance. Similar trends were observed on the ADReSS dataset, where \textbf{Cogno\_Z\_MEM} achieved the largest improvement, increasing performance from $F_1=0.81$ to $0.86$ ($\Delta F_1=0.05$).

The strongest evidence of external validity was observed on the PROCESS-2 benchmark. In the binary classification setting, all Cogno-derived strategies outperformed the original labels, with \textbf{Cogno\_Z\_MEM} producing the largest gain, increasing performance from $F_1=0.77$ to $0.85$ ($\Delta F_1=0.093$). Similar improvements were observed for Cogno\_G\_MEM ($\Delta F_1=0.063$), whereas FLAME-based attention and executive-function relabelling reduced performance substantially. In the multiclass setting, memory-based relabelling again produced the largest improvements, with \textbf{Cogno\_Z\_MEM} increasing performance from $F_1=0.51$ to $0.60$ ($\Delta F_1=0.090$), followed by Cogno\_G\_MEM ($\Delta F_1=0.057$). In contrast, FLAME-derived attention and executive-function strategies consistently degraded multiclass performance.

A consistent pattern emerged across all external datasets. Relabelling strategies derived from memory-domain neuropsychological performance yielded the largest and most stable improvements, whereas attention- and executive-function-based strategies produced smaller gains and FLAME-derived attention/executive thresholds frequently reduced performance. Notably, the normative z-score approach (Cogno\_Z) demonstrated the strongest cross-cohort transferability, achieving the highest performance improvements on Pitt, ADReSS, and PROCESS-2. These findings indicate that neuropsychological label refinement improves not only internal classification performance but also generalisation to independent datasets collected under substantially different conditions, supporting the hypothesis that the reassigned labels more closely approximate latent clinical ground truth than the original diagnostic annotations.

\subsection{Summary of performance improvement}

The performance summary is demonstrated in Table \ref{tab:combined_results}, and across both classification settings, memory-based labelling strategies consistently outperformed the original diagnostic labels.

In the internal binary classification setting, all memory-based methods (FLAME\_MEM, Cogno\_G\_MEM, and Cogno\_Z\_MEM) achieved large and statistically significant improvements over the original labels (all $p < 0.001$, Holm--Bonferroni corrected), with very large effect sizes (Cohen's $d > 2.6$). Among these, Cogno\_Z\_MEM achieved the highest improvement ($\Delta F_1 = 0.083$, $d = 3.31$), followed by Cogno\_G\_MEM and FLAME\_MEM.

Attention- and executive-based methods showed moderate improvements, particularly under Cogno\_Z and Cogno\_G formulations, although their effect sizes were smaller. In contrast, FLAME\_ATTEN significantly decreased performance ($\Delta F_1 = -0.032$, $p = 0.008$), indicating that not all domain-specific transformations are beneficial.

In the internal multiclass classification setting, performance gains were smaller but remained consistent. Memory-based methods again yielded the strongest improvements, with FLAME\_MEM ($\Delta F_1 = 0.061$, $d = 3.11$), Cogno\_Z\_MEM ($\Delta F_1 = 0.069$, $d = 2.82$), and Cogno\_G\_MEM ($\Delta F_1 = 0.053$, $d = 2.02$) all achieving statistically significant improvements ($p < 0.001$). Executive- and attention-based methods produced smaller but still significant gains under selected Cogno\_Z and Cogno\_G formulations, whereas the corresponding FLAME-based variants did not yield significant improvements.

External validation across three independent cohorts demonstrated similar trends. On the Pitt and \adress{} datasets, memory-based calibration consistently produced the largest improvements, with Cogno\_Z\_MEM achieving gains of $\Delta F_1 = 0.040$ and $\Delta F_1 = 0.050$, respectively. Likewise, on the PROCESS-2 benchmark, Cogno\_Z\_MEM yielded the strongest performance improvements in both binary ($\Delta F_1 = 0.093$) and multiclass ($\Delta F_1 = 0.090$) classification tasks. Although inferential statistics were not computed for the external datasets because of the limited number of available benchmark tasks, the consistent superiority of memory-derived relabelling across all three cohorts supports the generalisability of the proposed framework.

Overall, these results demonstrate that memory-derived labelling strategies are the most robust and effective across both internal and external validation settings. In particular, the normative Cogno\_Z\_MEM formulation consistently achieved the largest performance gains, suggesting that memory-centred neuropsychological calibration provides a reliable mechanism for reducing diagnostic label noise and improving automated cognitive assessment.

\begin{figure*}[!ht]
\centering 
\includegraphics[width=\linewidth]{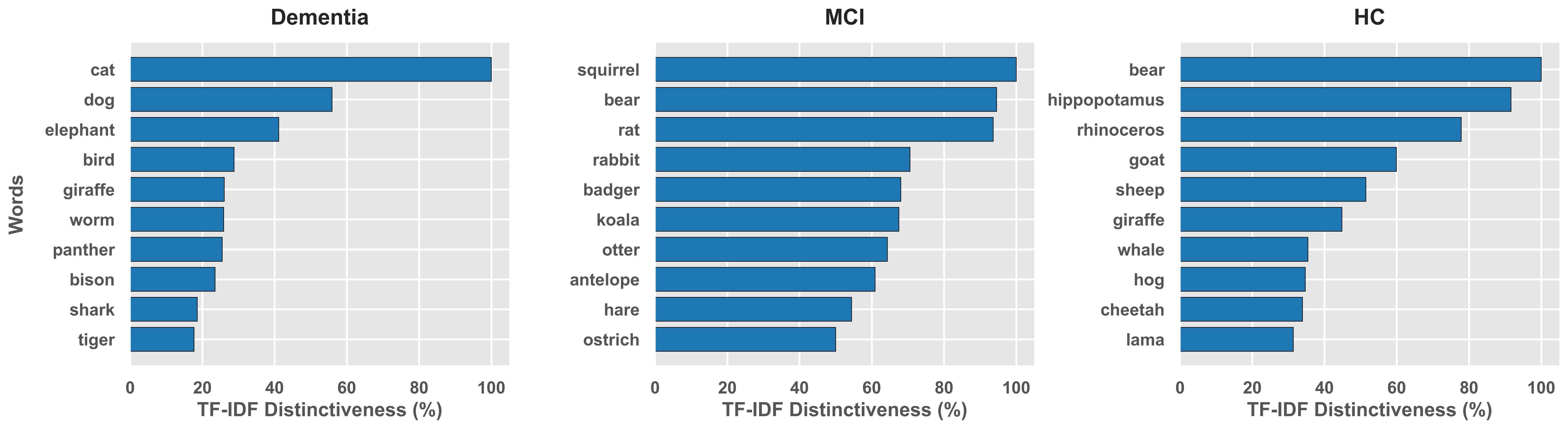} 
\caption{
\textbf{\tfidf{} analysis of semantic fluency responses using Cogno\_Z\_MEM diagnostic labels.}
The figure shows the ten most distinctive animal words produced by participants in each diagnostic group (Dementia, MCI, and Healthy Control).
\tfidf{} scores were normalised within each group and expressed as percentages relative to the most distinctive word (100\%) in that group.
Higher percentages indicate words that are disproportionately characteristic of a specific cognitive group rather than frequently used across all participants.
Distinct lexical patterns emerge across diagnostic categories, suggesting systematic differences in semantic retrieval strategies associated with cognitive impairment.
}
\label{fig:TF_IDF}
\end{figure*}


\section{Discussion}
\label{sec:discussion}

Across both binary and multiclass classification settings, neuropsychologically informed label generation substantially improved speech-based detection of cognitive impairment relative to original clinical diagnostic labels. Memory-derived calibration strategies consistently produced the strongest gains, increasing overall performance from $F_1 = 0.72$ to $0.83$ in the binary task and from $0.60$ to $0.68$ in the more challenging multiclass setting, for internal validation test set. 
Improvements of 4\%, 5\% and 9\% are also observed for the same memory-based calibration while tested on external Pitt, \adress{} and PROCESS-2 data sets. 
Improvements were observed across nearly all speech tasks and modelling configurations, indicating that performance gains arose primarily from enhanced label quality rather than changes in model architecture or transcription system. Confusion-matrix analyses further demonstrated reduced diagnostic ambiguity and improved sensitivity to impaired participants, particularly for cognitively demanding memory elicitation tasks, supporting memory impairment as the most reliable speech-derived marker of cognitive decline.

Incorporating global cognitive screening information via MoCA-derived priors produced additional systematic benefits, yielding an average $\sim$9--10\% relative improvement in $F_1$-score with extremely large effect sizes across experiments. These findings suggest that clinically anchored thresholding stabilises automatically generated labels and improves separation between dementia, mild cognitive impairment (MCI), and healthy ageing. Notably, gains persisted in the multiclass setting, where intermediate cognitive states are typically difficult to distinguish, indicating that label refinement rather than model complexity represents a key bottleneck in speech-based cognitive assessment. Collectively, the results demonstrate that neuropsychologically grounded relabeling, particularly memory-based strategies augmented by clinical priors, provides a robust pathway toward more reliable and clinically interpretable automated cognitive screening.

To investigate qualitative linguistic differences underlying classification performance, we performed a \tfidf{} analysis of semantic fluency responses using the \CognoZ\_MEM diagnostic labels (Fig.~\ref{fig:TF_IDF}). 
For each diagnostic group, \tfidf{} values were normalised relative to the most distinctive word within that group, allowing comparison of relative lexical specificity rather than absolute frequency.

Distinct semantic patterns emerged across cognitive states.
Participants classified as healthy controls produced a wider range of large or less prototypical animals (e.g., \textit{hippopotamus}, \textit{rhinoceros}, and \textit{whale}), suggesting preserved semantic search breadth and access to lower-frequency lexical items.
In contrast, dementia participants showed higher distinctiveness for more common or visually salient animals such as \textit{cat}, \textit{dog}, and \textit{bird}, consistent with restricted semantic retrieval and reliance on highly accessible category exemplars \cite{henry2004verbal, hodges1992semantic, monsch1992comparisons}.
The MCI group exhibited an intermediate pattern, characterised by mid-frequency animals such as \textit{squirrel}, \textit{rabbit}, and \textit{bear}, reflecting partial preservation of semantic diversity \cite{henry2004verbal}.

Although percentage values were not explicitly displayed in the figure, normalisation revealed that the most distinctive words within each group reached 100\% relative \tfidf{} importance, with subsequent words gradually decreasing in distinctiveness.
This gradient indicates that cognitive decline is associated not only with reduced lexical quantity but also with altered semantic organisation and category exploration strategies.

These findings support the hypothesis that neuropsychologically informed labelling captures meaningful linguistic structure rather than merely improving classifier optimisation, providing interpretability linking speech-derived biomarkers to underlying cognitive decline processes.

Importantly, these semantic differences emerge without manual linguistic annotation, demonstrating that neuropsychologically calibrated labelling enables automated discovery of clinically interpretable language biomarkers.


\subsection{Limitations}
\label{subsec:limitations}

Several limitations should be considered when interpreting the present findings. First, although neuropsychological-performance-based relabelling improved agreement with clinically verified diagnoses, the reassigned labels remain indirect proxies for formal clinical diagnosis and cannot replace comprehensive neurological assessment.

Second, neuropsychological thresholds were derived from available cohort distributions and may reflect demographic, educational, cultural, or recruitment biases inherent to the source population. Although the proposed framework generalised across three independent external datasets (Pitt, ADReSS, and PROCESS-2), further validation across more diverse international cohorts will be required to establish broader population-level applicability.

Third, the training cohort remained diagnostically imbalanced, with substantially fewer dementia participants than healthy controls. While the observed performance improvements were replicated across external datasets with differing class distributions, residual imbalance may still influence learned decision boundaries and performance estimates. Larger clinically verified cohorts with more balanced diagnostic representation would help further improve robustness.

Fourth, dementia participants were older on average than MCI and healthy-control participants, introducing the possibility of age-related linguistic and acoustic confounding. The consistent improvements observed across external datasets with different demographic characteristics suggest that age alone is unlikely to explain the reported gains; however, future work should explicitly model age, education, and other demographic factors to disentangle disease-related and demographic effects.

Fifth, multilingual participants were excluded because current ASR and language models remain insufficiently reliable for multilingual cognitive assessment. Consequently, the findings presently generalise primarily to monolingual English-speaking populations.

Sixth, speech-derived biomarkers are influenced by factors unrelated to neurodegeneration, including recording conditions, microphone characteristics, speaker variability, and emotional state, which may introduce residual noise despite improved labelling strategies.

Finally, although multiple ASR and language models were evaluated, the study did not exhaustively explore all possible architectures. Nevertheless, the consistent gains observed across model configurations and external datasets suggest that label quality exerts a larger influence on performance than model selection alone. The framework was additionally validated retrospectively, and prospective deployment studies will be required to determine whether neuropsychologically anchored relabelling improves real-world screening accuracy, clinical workflow integration, and longitudinal disease monitoring.


\section{Conclusion}
\label{sec:conclusion}

This study introduced a diagnostic label reassignment framework based on stratified neuropsychological performance for improving automatic detection of cognitive decline from real-world conversational speech collected through the \cognospeak{} platform. 
Rather than treating diagnostic annotations as fixed supervision, we reconstructed diagnostic labels using objective neuropsychological-performance measures across memory, attention, and executive function domains, enabling robust model training under uncertain or weakly supervised clinical annotations.

Cross-dataset comparison between \cognospeak{} and the FLAME cohort demonstrated broadly consistent neuropsychological trends despite observable scale differences. While attention and executive function scores showed greater inter-dataset variability, memory performance remained strongly aligned across cohorts, supporting the external validity and generalisability of neuropsychologically grounded representations.

Within the \cognospeak{} dataset, comparisons between clinical and non-clinical recruitment sources revealed domain-dependent differences. Attention and executive function measures exhibited occasional shifts attributable to acquisition context, whereas memory performance remained stable across sources. This robustness indicates that memory-derived neuropsychological signals are less sensitive to data provenance and, therefore, particularly suitable for large-scale semi-automatic dataset curation.

Across both binary and multiclass classification tasks, neuropsychologically informed relabeling consistently improved performance relative to original diagnostic annotations. Memory-based strategies achieved the highest gains, with the Cogno\_Z\_MEM approach emerging as the strongest overall performer. Binary classification performance increased by 13\% to an $F_1$-score of 0.83, while multiclass performance reached 0.66 with a 12\% increase despite the increased difficulty of separating intermediate cognitive states such as MCI. 
Improvements of 4\%, 5\% and 9\% are also observed for the same memory-based calibration while tested on external Pitt, \adress{} and PROCESS-2 data sets. 
Notably, incorporating thresholds derived from a relatively small number of clinically validated observations substantially improved performance across experiments, demonstrating that weakly supervised learning benefits strongly from neuropsychological performance-based decision boundaries.

Beyond quantitative improvements, \tfidf{} analysis revealed systematic semantic differences across diagnostic groups. Dementia participants relied predominantly on highly prototypical category exemplars, whereas healthy controls exhibited broader semantic exploration. These findings indicate that neuropsychologically informed labelling uncovers clinically interpretable linguistic biomarkers rather than merely improving classifier optimisation.

Overall, the results demonstrate that accuracy of diagnostic labels represents a critical bottleneck in speech-based cognitive assessment. By integrating stratified neuropsychological performance with weak-supervision machine learning, the proposed framework enables scalable, clinically interpretable, and remotely deployable cognitive screening using conversational speech.

Future work will investigate adaptive threshold learning, multimodal biomarker fusion, cross-cohort validation, and strategies to address remaining dataset imbalances, while incorporating additional cognitive screening instruments, including the Rowland Universal Dementia Assessment Scale (RUDAS), Addenbrooke's Cognitive Examination (ACE)-III, and Multicultural Cognitive Examination (MCE), alongside MoCA to further strengthen neuropsychologically grounded diagnostic calibration and clinical translation.

\section{Acknowledgments}

This research was partly funded by the NIHR Sheffield Biomedical Research Centre (BRC), and the NIHR202911 award under the NIHR i4i programme. The views expressed are those of the authors and not necessarily those of the NHS, the NIHR or the Department of Health and Social Care (DHSC).
Ethical approval for the collection of the data analysed in this study was granted by the NRES Committee South West-Central Bristol (REC number 16/LO/0737).
We thank Professors Anne Corbett and Clive Ballard for their valuable advice regarding the FLAME and PROTECT studies. 
For the purpose of open access, the author has applied a Creative Commons Attribution (CC BY) licence to any Author Accepted Manuscript version arising. 
All participants took part in this study voluntarily and provided their consent to use their data for research purposes.

\section{Data Availability}

The neuropsychological performances along with demographic information for the available 720 participants are shared in the GitHub repository; however due to ethical constraints, no audio has been shared.

\begin{center}
\url{https://github.com/Madhurananda/CognoCalib}
\end{center}

\section{Code Availability}

All code required to reproduce statistical analyses, embedding 
generation, and baseline modelling experiments along with associated Anaconda environments are publicly available at:

\begin{center}
\url{https://github.com/Madhurananda/CognoCalib}
\end{center}

The codebase is released under the Apache License 2.0, permitting reuse, 
modification, and redistribution subject to the terms of the licence. 
Version control is maintained through GitHub to ensure transparency and 
reproducibility of all results reported in this study.
Code archived at Zenodo DOI \cite{madhurananda_pahar_2026_23263139}:
\begin{center}
\url{https://doi.org/10.5281/zenodo.23263139}
\end{center}



\bibliographystyle{IEEEtran}
\bibliography{mybib}

\section{Author Contributions}
Madhurananda Pahar conceived the study, designed and conducted the experiments, performed data analysis, classification, and authored the published Python code and manuscript. 
Annalena Venneri supervised the MCI sorting process using statistical and hypothesis testing. 
Adam Hampshire, Peter J Hellyer, and Agata J Hampshire are the original authors of the Cognitron study. 
Caitlin Illingworth and Daniel Blackburn led participant recruitment and data collection across the community centres and assisted in writing the Methods section. 
Heidi Christensen supervised the research, contributed to the study design and interpretation of results, and provided overall project guidance.

All authors contributed to reviewing and editing the manuscript and approved the final version.

\section{Competing Interests}
The authors declare no competing interests.

\section{Funding}
This research was partly funded by the NIHR Sheffield Biomedical Research Centre (BRC), and the NIHR202911 award under the NIHR i4i programme.

\end{document}